\documentclass[preprint2]{aastex}

\def\bum{$\beta$-UMi}
\def\hwpss{HWP synchronous signal}
\def\ss{synchronous signal}

\usepackage{color}

\title{MAXIPOL: Cosmic Microwave Background Polarimetry \\ Using a Rotating Half-Wave Plate}

\author{B.~R.~Johnson\altaffilmark{5}, J.~Collins\altaffilmark{2},
M.~E.~Abroe\altaffilmark{1}, P.~A.~R.~Ade\altaffilmark{3}, J.~Bock\altaffilmark{4}, 
J.~Borrill\altaffilmark{7,9}, A.~Boscaleri\altaffilmark{13} , P.~de~Bernardis\altaffilmark{12}, 
S.~Hanany\altaffilmark{1}, A.~H.~Jaffe\altaffilmark{8}, T.~Jones\altaffilmark{1}, 
A.~T.~Lee\altaffilmark{2,6,9}, L.~Levinson\altaffilmark{10}, T.~Matsumura\altaffilmark{1}, 
B.~Rabii\altaffilmark{2}, T.~Renbarger\altaffilmark{1}, P.~L.~Richards\altaffilmark{2}, 
G.~F.~Smoot\altaffilmark{2,6,9}, R.~Stompor\altaffilmark{14}, H.~T.~Tran\altaffilmark{2,9}, 
C.~D.~Winant\altaffilmark{2}, J.~H.~P.~Wu\altaffilmark{11}, J.~Zuntz\altaffilmark{8}}

\altaffiltext{1}{School of Physics and Astronomy, University of Minnesota, Minneapolis, MN, 55455, USA}
\altaffiltext{2}{Department of Physics, University of California, Berkeley, CA, 94720, USA}
\altaffiltext{3}{School of Physics and Astronomy, Cardiff University, Cardiff, UK, CF24 3YB}
\altaffiltext{4}{Jet Propulsion Laboratory, Pasadena, CA, 91109, USA}
\altaffiltext{5}{Astrophysics, University of Oxford, Oxford, UK, OX1 3RH}
\altaffiltext{6}{Physics Division, Lawrence Berkeley National Lab, Berkeley, CA, 94720, USA}
\altaffiltext{7}{Computational Research Division, Lawrence Berkeley National Lab, Berkeley, CA, 94720, USA}
\altaffiltext{8}{Astrophysics Group, Blackett Lab, Imperial College, London, UK, SW7 2AZ}
\altaffiltext{9}{Space Sciences Laboratory, University of California, Berkeley, CA, 94720, USA}
\altaffiltext{10}{Department of Particle Physics, Weizmann Institute of Science, Rehovot, Israel}
\altaffiltext{11}{Department of Physics, Institute of Astrophysics, \& Center for Theoretical Sciences, 
                  National Taiwan University, Taipei 10617, Taiwan}
\altaffiltext{12}{Dipartimento di Fisica, Universita di Roma La Sapienza, Italy}
\altaffiltext{13}{IFAC-CNR, Firenze, Italy}
\altaffiltext{14}{Laboratoire AstroParticule et Cosmologie, Universit{\'e} Paris-7, Paris, France}

\begin{abstract}

We discuss MAXIPOL, a bolometric balloon-borne experiment designed to
measure the E-mode polarization of the cosmic microwave background
radiation (CMB).  MAXIPOL is the first bolometric CMB experiment to
observe the sky using rapid polarization modulation.  To build
MAXIPOL, the CMB temperature anisotropy experiment MAXIMA was
retrofitted with a rotating half-wave plate and a stationary analyzer.
We describe the instrument, the observations, the calibration and the
reduction of data collected with twelve polarimeters operating at
140~GHz and with a FWHM beam size of 10~arcmin. We present maps of the
$Q$ and $U$ Stokes parameters of an 8~deg$^2$ region of the sky near
the star \bum.  The power spectra computed from these maps give weak
evidence for an $EE$ signal.  The maximum-likelihood amplitude of
$\ell(\ell+1)C^{EE}_{\ell}/2\pi$ is $55_{-45}^{+51}~\mu\mbox{K}^2$
(68\%), and the likelihood function is asymmetric and skewed positive
such that with a uniform prior the probability that the amplitude is
positive is 96\%.  This result is consistent with the expected
concordance $\Lambda$CDM amplitude of $14~\mu\mbox{K}^2$. The maximum
likelihood amplitudes for $\ell(\ell+1)C^{BB}_{\ell}/2\pi$ and
$\ell(\ell+1)C^{EB}_{\ell}/2\pi$ are $-31_{-19}^{+31}$ and
$18_{-34}^{+27}~\mu\mbox{K}^2$ (68\%), respectively, which are
consistent with zero.  All of the results are for one bin in the range
$151 \leq \ell \leq 693$.  Tests revealed no residual systematic
errors in the time or map domain.  A comprehensive discussion of the
analysis of the data is presented in a companion paper.

\end{abstract}


\begin{document}

\maketitle


\section{Introduction}
\label{sec:introduction}

\begin{figure*}[ht]
\begin{center}
\includegraphics[width=5.in]{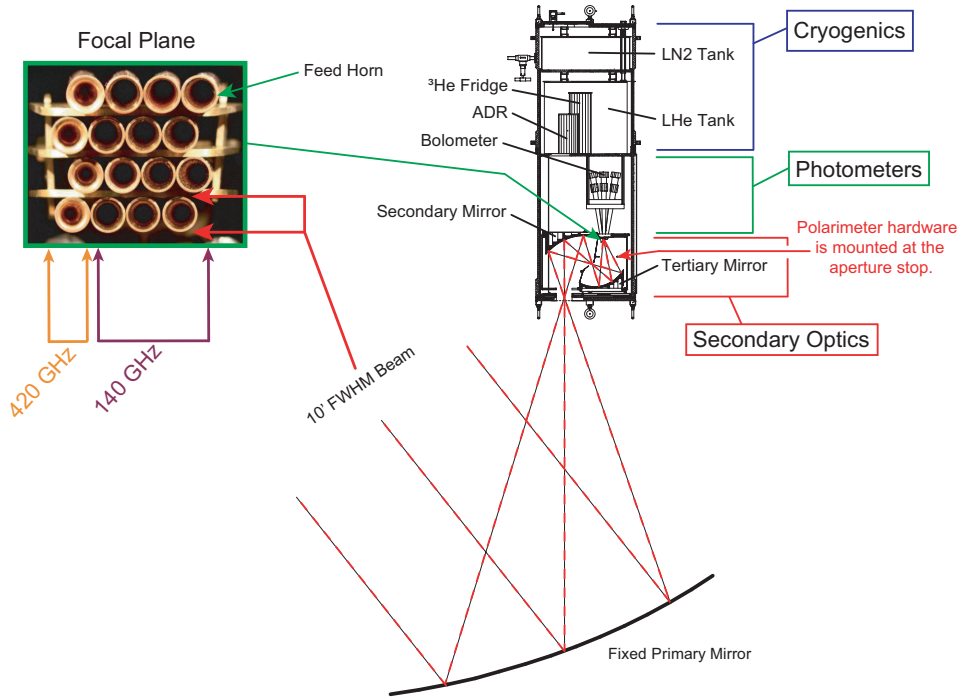}
\caption{\footnotesize{An overall view of the MAXIPOL instrument.
    Most of the elements shown here were shared with the previous
    MAXIMA experiment. The polarimeter hardware, which is not shown
    here, is shown in Figure~\ref{fig:hwp}.}}
\label{fig:cross}
\end{center}
\end{figure*}

Measurements of the polarization of the cosmic microwave background
radiation (CMB) can confirm fundamental predictions made by our
cosmological model and probe the period after the Big Bang when
inflation is believed to have occurred. Several experiments recently
reported detection of either the $TE$ or $EE$ power spectrum
\citep{cbi, capmap, boomerang, dasi, wmap} and there is on-going
effort to build even more sensitive receivers that can probe the small
signal amplitudes that are predicted for the $BB$ power spectrum.

MAXIPOL is a bolometric, balloon-borne experiment designed to measure
the polarization of the CMB. It uses a rotating half-wave plate (HWP)
and a wire grid analyzer to modulate signals reaching the bolometers
from polarized sources.  MAXIPOL is unique in that it is the only
bolometric CMB experiment to date to deploy rapid polarization
modulation. With our technique, the polarized component of the
incident radiation is modulated at a frequency equal to four times the
rotation frequency of the HWP.  This property is advantageous for
measuring the small CMB polarization because polarized sky signals are
moved to a narrow band in the frequency domain which can be far from
detector 1/$f$ noise or other spurious instrumental signals. In
particular, there is no confusion with spurious signals that appear at
the modulation frequency itself.  Because only polarized signals are
modulated, a HWP polarimeter separates the temperature and
polarization signals in the frequency domain, making analysis of the
signals independent.  Of particular importance is the property that
each detector is an independent polarimeter measuring the Stokes
parameters $I$, $Q$, and $U$ over a relatively short time.  Several
systematic errors that complicate detector differencing techniques are
avoided.

Differencing polarimeters are defined here as instruments that measure
Stokes parameters by differencing the signal from two detectors, each
of which is sensitive to one of the orthogonal linear polarizations.
Differencing polarimeters can be constructed using individual
polarizers placed at the entrance aperture of each photometer, two
bolometers with orthogonal absorbing grids, or orthomode transducers
which split the incoming polarization into two components \citep[see
for example][]{boom_inst,Archeops}.  Differencing polarimeters only
modulate polarized sky signals through sky rotation and telescope
scanning.  They are prone to spurious polarization signals through
errors in the absolute calibration of detector pairs, time dependent
responsivity variations, noise properties that are not common to both
detectors, or different antenna patterns for the two detectors.
Polarization modulators like the one used in MAXIPOL can mitigate
these problems.

Polarization modulation technologies different from MAXIPOL's include
rotating polarizers, photoelastic modulators and Faraday rotation
modulators.  Rotating polarizers reflect the unwanted polarization
component and can thus produce spurious signals through multiple
reflections inside the receiver.  Suitable photoelastic modulators
have not been developed in the frequency bands of interest. Broadband
Faraday rotation modulators for the millimeter-wave band are just now
being developed \citep{bicep} and little is known about systematic
errors associated with their operation. Modulation of sky signals with
a rotating HWP and stationary analyzer is a proven technique in
infrared and millimeter-wave astrophysics \citep{tinbergen} and
therefore we chose to implement the technique for CMB polarimetry.
Since MAXIPOL is the first CMB experiment to produce results using
this strategy, the experience gained from hardware implementation,
data analysis and characterization of HWP-specific systematic errors
will inform the design of next-generation experiments that aim to
characterize the anticipated B-mode signals.

To build MAXIPOL, the receiver from the CMB temperature anisotropy
experiment MAXIMA \citep{hanany,balbi,lee,stompor,abroe} was converted
into a polarimeter by retrofitting it with a HWP and a fixed
polarization analyzer.  The rest of the MAXIMA instrument including
the detector system, cryogenics, optics, and electronics was
essentially unchanged.  MAXIPOL observed the sky during two flights
that were launched from the NASA Columbia Scientific Ballooning
Facility (CSBF) in Ft.~Sumner, New Mexico. The first flight, launched
in 2002 September, yielded less than an hour of useful data because of
a telemetry failure.  The primary CMB data set, which will be
discussed in this paper, was collected during the second flight in
2003 May.

We describe the instrument and the observations in
Sections~\ref{sec:instrument}~and~\ref{sec:observations}.  The
HWP-specific information appears in
Sections~\ref{sec:pointing},~\ref{sec:tod},~\ref{sec:calibration},
and~\ref{sec:systematics}.  In particular, Section~\ref{sec:tod}
describes the processing of our time-domain data. This information
will be useful for future CMB experiments that will use HWP
polarimeters. Estimated power spectra and $Q$ and $U$ maps are
presented in a summary of our analysis in Section~\ref{sec:results}.
A comprehensive description of the analysis is given in \citet{proty}.


\begin{figure}[h]
\begin{center}
\includegraphics[width=2.75in]{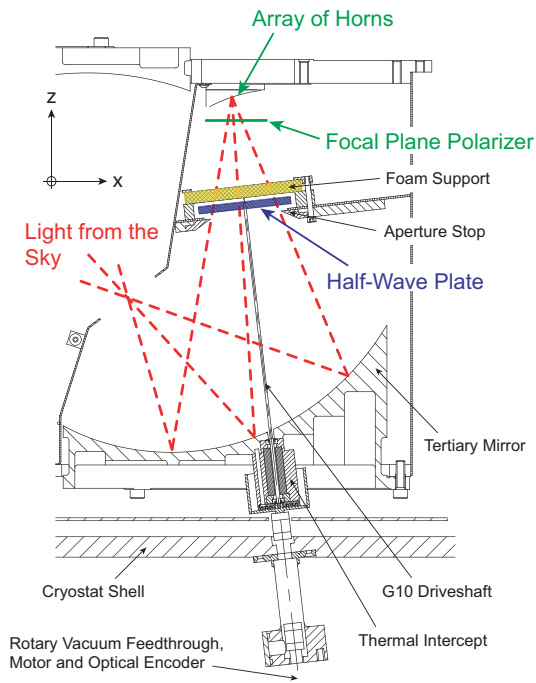} 
\caption{\footnotesize{A cross-sectional view of the receiver near the
focal plane showing the polarimetry components. The HWP was mounted at
an aperture stop of the optical system and a roof-shaped polarization
analyzer was mounted in front of the entrance apertures of the horns.
A sketch showing the analyzer and the focal plane projected on a
perpendicular plane is given in Figure~\ref{fig:focal_plane}.}}
\label{fig:hwp}
\end{center}
\end{figure}

\section{Instrument}
\label{sec:instrument}

In this Section, we give an overview of the experiment emphasizing the
polarimeter. Additional technical details are given in previous MAXIMA
publications
\citep{lee_hardware,hanany,rabii_phd,winant_phd,rabii_winant} and
MAXIPOL papers
\citep{johnson,johnson_phd,johnson_moriond,collins_phd}.

MAXIPOL employed a three-mirror telescope with a 1.3 m off-axis
parabolic primary mirror. The elliptical secondary and tertiary
reimaging mirrors were maintained at liquid helium temperatures inside
the receiver to reduce radiative loading on the bolometers. The
primary mirror, which was nutated to produce rapid azimuth modulation
for MAXIMA, was fixed for MAXIPOL to avoid modulating the polarization
properties of the telescope.

Millimeter-wave radiation from the sky was re-imaged to a $4 \times 4$
array of photometers at the focal plane.  Observations were made in
bands centered on 140 and 420~GHz with $\Delta\nu \simeq$ 30~GHz.  The
twelve 140~GHz photometers were optimized to measure the CMB and the
four 420~GHz photometers were used to monitor foreground dust
contamination. The 10$^\prime$ FWHM Gaussian beam shape for the
140~GHz photometers was defined by a smooth-walled, single-moded
conical horn and a cold Lyot stop. The 420~GHz photometers employed
multi-mode Winston horns.  The bolometers were maintained at 100~mK by
the combination of an adiabatic demagnetization refrigerator
\citep{hagman} and a 300~mK $^{3}$He refrigerator.  A photograph of
the focal plane and a cross-sectional overview of the optical system,
which was essentially unchanged from MAXIMA, are shown in
Figure~\ref{fig:cross}.

Figure~\ref{fig:hwp} shows a cross-section of the portion of the cold
optics that was modified to convert MAXIMA to MAXIPOL.  We used a
3.175~mm thick A-cut sapphire HWP. Reflections from the HWP were
minimized by bonding 330~$\mu$m thick wafers of Herasil to each face
of the sapphire with Eccobond 24, an unfilled, low viscosity epoxy
that was used to achieve glue layers as thin as 13~$\mu$m.  The HWP
thickness was selected to minimize the fraction of emerging
elliptically polarized intensity and thereby optimize the overall
modulation efficiency of the polarimeter.  The calculated efficiency
incorporated the effects of the finite spectral bandwidth of the
photometers and the oblique incidence of rays on the HWP.

Since the anti-reflection coating was not birefringent, the two
incident polarization orientations had different coefficients of
reflection.  This differential reflection gave rise to a HWP
synchronous signal at a frequency of $2f_{o}$, where $f_{o}$ is the
rotation frequency of the HWP.  To minimize this effect, we calculated
the anti-reflection coating thickness that would minimize the
difference in reflection coefficients given the spectral bandwidth of
the 140~GHz photometers, the thickness of the Eccobond 24 layer and
the angles of the rays.  The $2f_{o}$ signal was not a source of
systematic error because it was out of the polarization signal band
centered on $4f_{o}$ (see Section~\ref{sec:tod}).

\begin{figure}[h]
\begin{center}
\includegraphics[width=2.75in]{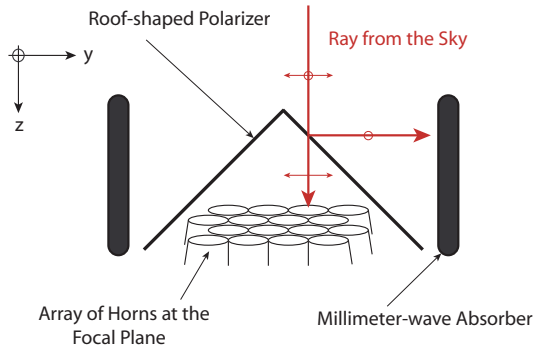}
\caption{\footnotesize{A sketch of the roof-shaped analyzer positioned
in front of the array of horns at the focal plane.}}
\label{fig:focal_plane}
\end{center}
\end{figure}

The polarization analyzer was a commercial grid polarizer epoxied to a
rigid ``roof-shaped'' frame that was positioned over the horn openings
(see Figure~\ref{fig:focal_plane}).  This roof-shaped polarizer was
implemented so that the light that was reflected by the grid was
directed out of the optical path and into a cold millimeter-wave
absorber \citep{bockblack} that was mounted on both sides of the focal
plane.  The grid polarizer had 98 electroformed gold stripes per cm
that were bonded onto a 38~$\mu$m thick Mylar substrate.  The stripes
were 5~$\mu$m wide.

The HWP was located near the Lyot stop and was driven from its axis at
$f_{o}$~=~1.86~Hz during all observations (see Figure~\ref{fig:hwp}).
The orientation of the HWP was measured with a 17-bit optical
encoder. This rotation frequency, combined with the azimuth scan
frequency that was either 0.10~Hz or 0.06~Hz during the primary CMB
observations, gave approximately three to five modulations of the $Q$
and $U$ Stokes parameters per 10~arcmin sky beam in one azimuthal
scan.


\section{Observations and Scans}
\label{sec:observations}

The instrument was launched at 15:14 UT (9:14 AM Local Time) on 2003
May 24.  The first observation began at 18:08 UT when the payload
reached an altitude of 38.7 km. The flight terminated 26 hours after
launch, 477~km west and 120~km south of the launch site.  In this
paper we only discuss the data collected during one 7.6 hour-long
nighttime CMB observation near Beta Ursae Minoris (\bum), RA = 14h
50$^{\prime}$ 42.5$^{\prime\prime}$, Dec= +74$^{\circ}$ 09$^{\prime}$
42.$^{\prime\prime}$ Other observations are discussed in
\citet{johnson_phd} and \citet{collins_phd}.

\begin{figure}[h]
\begin{center}
\includegraphics[width=3.in]{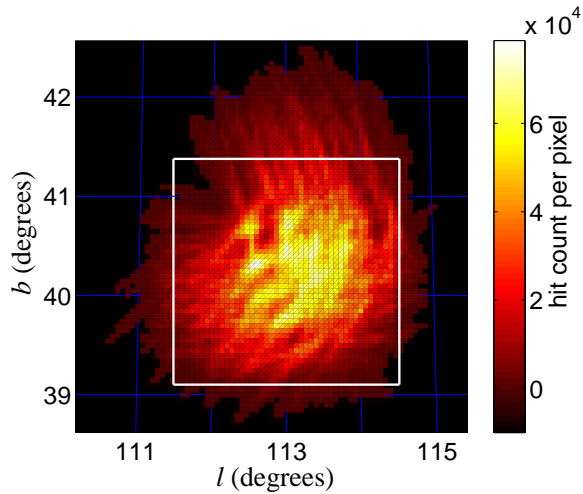}
\caption{\footnotesize{Map of the total number samples per
    3$^{\prime}$ pixel from the twelve 140~GHz polarimeters for the
    \bum\ observation. The white square marks the $2.3^{\circ} \times
    2.3^{\circ}$ region used for power spectrum estimation.}}
\label{fig:hits}
\end{center}
\end{figure}

Jupiter was observed during both the day and nighttime portions of
this flight to obtain the absolute intensity calibration of the
instrument, to map the telescope beam shapes, and to measure the
location of the beam centers for pointing reconstruction.  During
these observations the gondola was scanned in azimuth with a
triangle-wave profile centered on Jupiter.  The turnarounds on this
profile were smoothed to ensure quiescent noise performance from the
bolometers.  The daytime scans were $\pm 0.6^{\circ}$ with a 20~sec
period.  The nighttime scans were $\pm 1.3^{\circ}$ with an 18~sec
period.  Simultaneously, the telescope elevation was rastered $\pm
0.5^{\circ}$ relative to the planet every 20~min.

For the CMB observations the telescope azimuth was scanned in a
similar fashion near \bum, but with an amplitude of $1^{\circ}$ and a
period that was changed mid-scan from 10 to 15~sec.  To improve
cross-linking the telescope elevation was re-pointed $0.2^{\circ}$
above and below the guide star every 10~min.  Figure~\ref{fig:hits}
shows the total number of 140~GHz detector samples per 3~arcmin pixel
produced by this scan strategy.  Observation statistics including
pointing error for all scans are given in
Table~\ref{table:scan_stats}.

\begin{table*}[ht]
\centering
\begin{tabular}{cccccccc}
 & & & \multicolumn{2}{c}{Pointing Error} \\
Scan Target  & Start Time  & Duration & Random & Systematic \\
             & [UT] & [hours]  & [arcmin] & [arcmin] \\
\hline
Daytime Jupiter    & 20:03 & 1.02 & 0.99 & 0.31 \\
Nighttime Jupiter  & 02:42 & 0.52 & 0.47 & 0.21 \\
\bum\              & 03:18 & 7.64 & 0.32 & 0.27 \\
\hline
\end{tabular}
\caption{\footnotesize{Scan targets with scan durations and pointing
    error. A detailed description of the pointing errors is given in
    \citet{collins_phd}. }}
\label{table:scan_stats}
\end{table*}


\section{Attitude Reconstruction}
\label{sec:pointing}

Telescope pointing was primarily determined with one of two star
cameras. The camera used during daytime observations was filtered with
a 695~nm Schott glass filter to reduce the brightness of the daytime
sky, and fitted with a reflective lens that had a 500~mm focal length,
which provided a FOV of 0.72$^{\circ}$ by 0.55$^{\circ}$.  The
unfiltered nighttime camera used a 50~mm lens that provided a
7.17$^{\circ}$ by 5.50$^{\circ}$ FOV. Pointing reconstruction using
star camera data has been described in previous publications
\citep{hanany,rabii_phd,rabii_winant,collins_phd}.

MAXIPOL adopts the convention for the Stokes parameters $I$, $Q$, and
$U$ used by WMAP, which references the polarization direction to the
NGP \citep{WMAPdataanalysis,zaldarriaga97}.  When the instrument
observes pixel $p$ at time $t$, the output of the detectors (in units
of volts and ignoring noise and systematic errors that will be
discussed later) can be written
\begin{eqnarray}
d_{p}(t) = R \, [ \, I_{p} 
\!\!\!\! &+& \!\!\!\! \epsilon \, Q_{p} \, \cos \left( \, 2 \gamma(t) \, \right) \nonumber \\ 
\!\!\!\! &+& \!\!\!\! \epsilon \, U_{p} \, \sin \left( \, 2 \gamma(t) \, \right) \, ]. 
\label{eq:WMAP.timestream}
\end{eqnarray}
Here $\gamma$ is the angle between the polarization reference vector
at sky pixel $p$ and the polarimeter transmission axis, $\epsilon$ is
the polarimeter modulation efficiency, and $R$ is an overall
calibration factor that has units of
V~K$^{-1}$~\citep{collins_phd}. The angle $\gamma$ is given by
\begin{equation}
\gamma = \alpha - 2\beta - \frac{\pi}{2}.
\label{eq:Beam_Angle1}
\end{equation}
Here, $\alpha$ is the angle between the reference vector at pixel $p$
and a vector pointing from $p$ to the zenith along a great circle,
$\beta$ is the HWP orientation angle in the instrument frame, and the
$\pi/2$ accounts for the fact that the transmission axis of the
analyzer was perpendicular to the zenith axis.  The angle $\alpha$ was
computed for each time sample from the pointing reconstruction.  The
angle $\beta$ was computed by subtracting a photometer-dependent
offset from the HWP encoder angle.  This encoder offset was measured
during laboratory calibration (see
Section~\ref{sec:polarimeter_characterization}).  By combining
Equations~\ref{eq:WMAP.timestream}~and~\ref{eq:Beam_Angle1} the model
for noiseless time ordered data (TOD) can be written
\begin{eqnarray} 
d(t) = R \, [ \, I_p 
\!\!\!\! &-& \!\!\!\! \epsilon \, Q_p \, \cos( \, 4 \beta(t) - 2 \alpha(t) \, ) \nonumber \\
\!\!\!\! &+& \!\!\!\! \epsilon \, U_p \, \sin( \, 4 \beta(t) - 2 \alpha(t) \, ) \, ].
\label{eq:final.timestream}
\end{eqnarray}


\section{Time Ordered Data Processing}
\label{sec:tod}

\begin{figure}[h]
\vspace{0.125in}
\begin{center}
\includegraphics[width=2.85in]{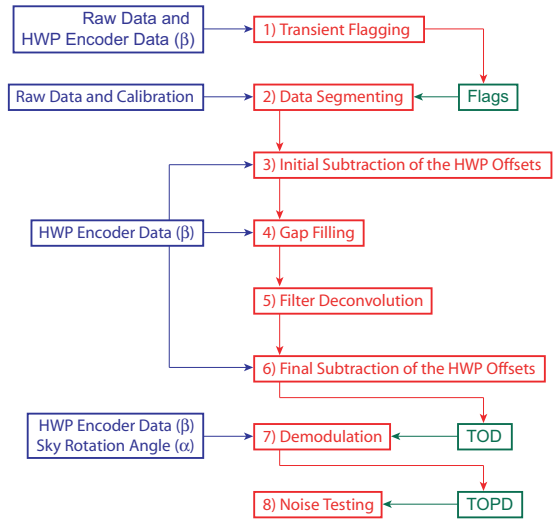}
\end{center}
\caption{\footnotesize{A flowchart of the time-domain processing
    algorithm. Blue boxes at left are data inputs, numbered red boxes
    in the middle are analysis operations and green boxes at right are
    data outputs.}}
\label{fig:TODFlowchart}
\end{figure}

In this paper we report on the data from the twelve 140~GHz
polarimeters.  The raw data from \bum\ consists of 5.76 million
time-ordered samples for each of the 140~GHz polarimeters. The
bolometer sample period was 4.8~msec.  We first give an overview of
the time-domain data-processing algorithm, which is outlined in
Figure~\ref{fig:TODFlowchart}, and then discuss some of the processing
steps in more detail. The rest of the data analysis procedure is
described in the companion paper, \citet{proty}.

Transients such as cosmic ray hits and internal calibration pulses
were flagged using an algorithm described in \citet{johnson_phd}.  The
data were calibrated and blocks of data separated by gaps longer than
30~sec were processed as separate segments.  These time domain data
contained a significant instrumental signal that was synchronous with
the rotation of the HWP.  We refer to this instrumental signal as the
\hwpss\ (or sometimes in short the \ss).  An initial estimate of the
\ss\ was subtracted. Gaps shorter than 30~sec were filled with
constrained noise realizations, and the \ss\ was replaced, leaving
continuous, transient-free data.  The instrumental filters were
deconvolved after padding the ends of each segment with 100~msec of
matched white noise, simulated \hwpss, and a window function that
smoothly decayed to zero at the endpoints. A section of data one
filter-time-width long was removed from the ends of each data segment
after filter deconvolution to avoid any contamination from edge
effects produced by the Fourier transforms.  The \hwpss\ was then
re-estimated and subtracted from the raw data, producing the
time-ordered data (TOD). The time-ordered polarization data (TOPD)
were extracted from the TOD by demodulation using a phase-locked,
sine-wave reference signal constructed from the angle $2 \gamma$.  The
TOPD were checked for noise stationarity using a frequency-domain
$\chi^2$ test and for Gaussianity using the Kolmogorov-Smirnoff (KS)
test in the time domain.

The following subsections describe the details of the \hwpss\
estimation, the filter deconvolution, the TOPD demodulation, and the
tests of the noise properties.


\subsection{Subtraction of the HWP Synchronous Signal}
\label{sec:hwpss_subtraction}

\begin{figure*}[ht]
\begin{center}
$
\begin{array}{cc}
\includegraphics[height=2.6in,angle=90]{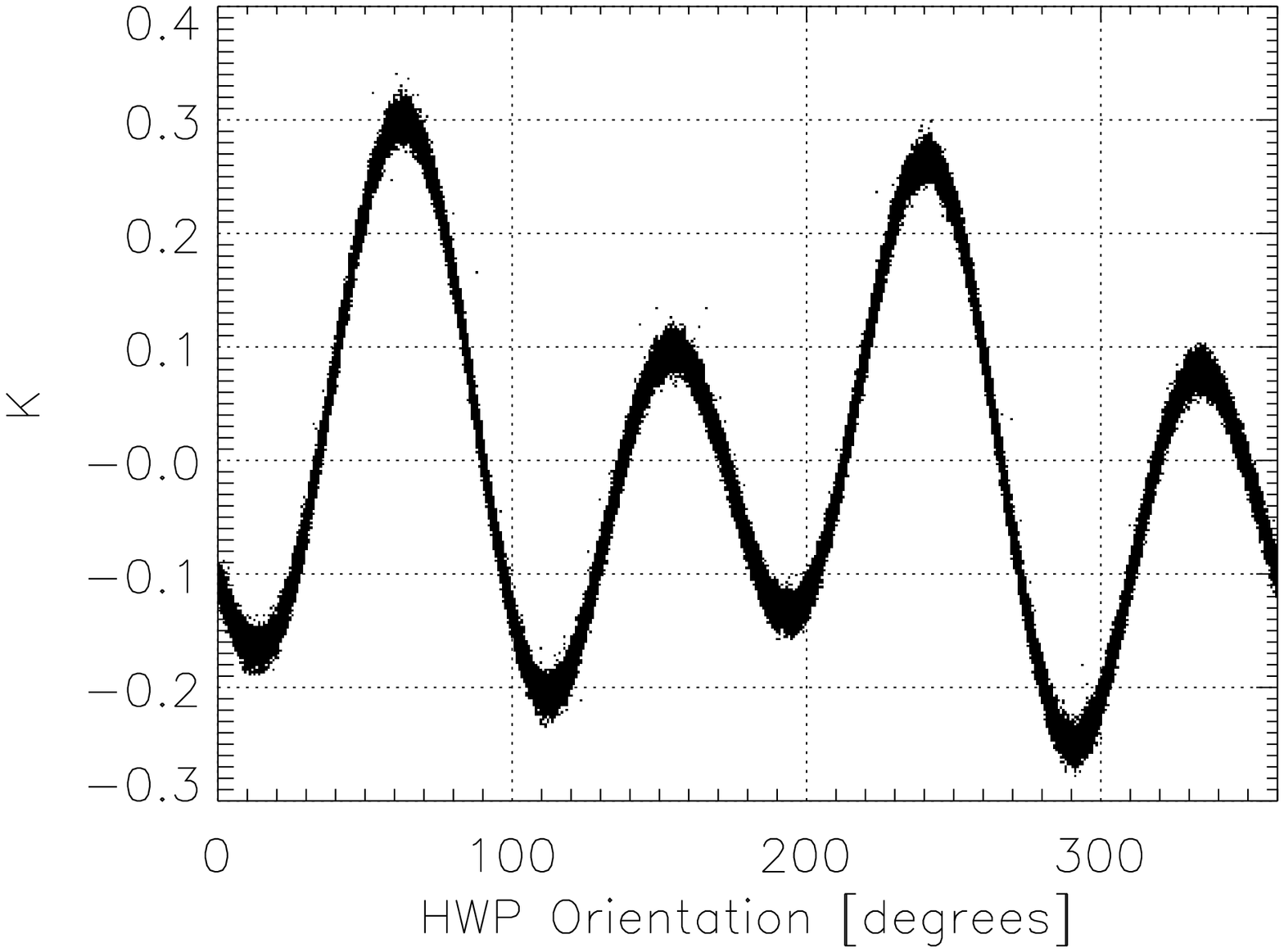} &
\includegraphics[height=2.6in,angle=90]{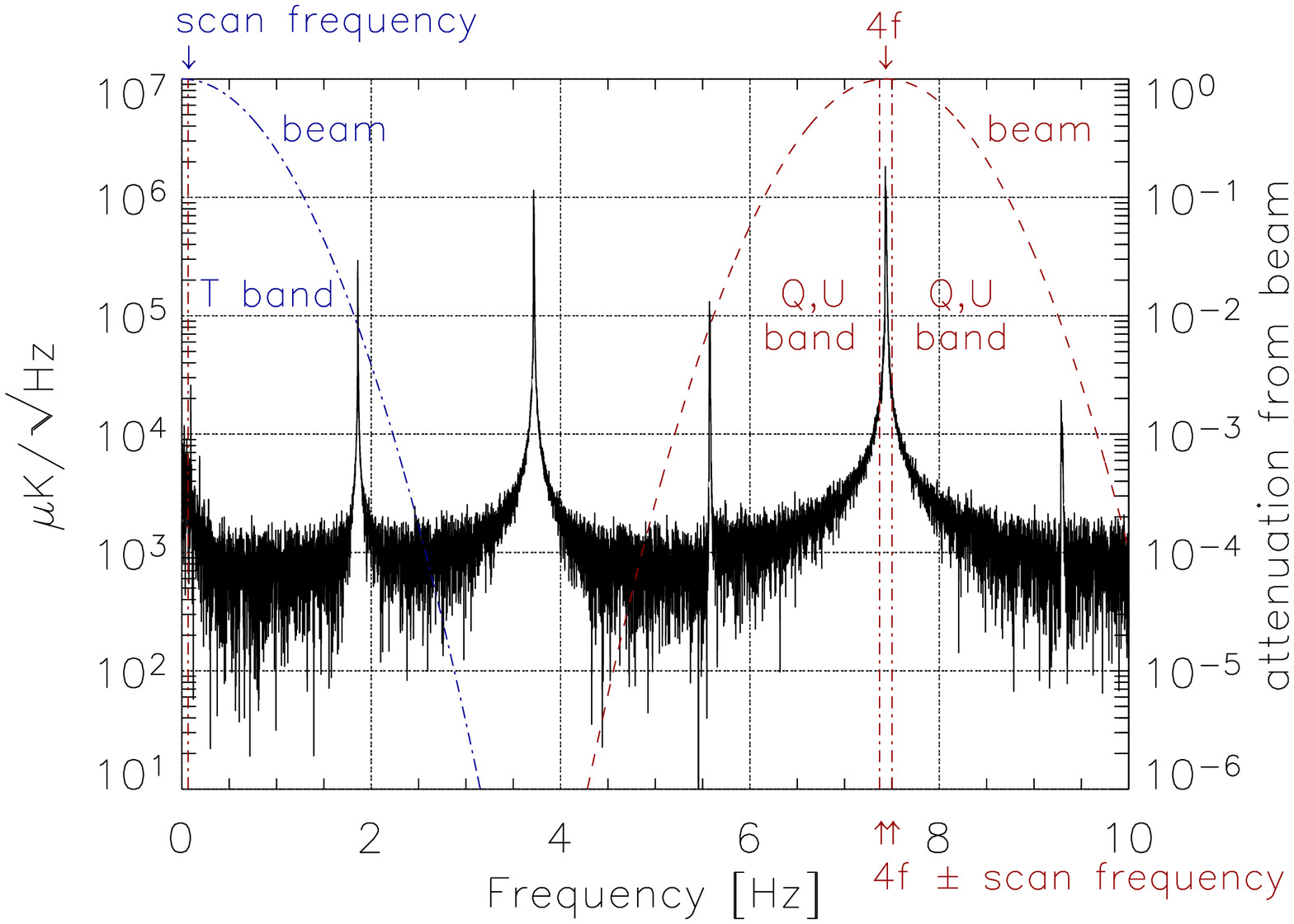} \\
(a) & (b) \\
\includegraphics[height=2.6in,angle=90]{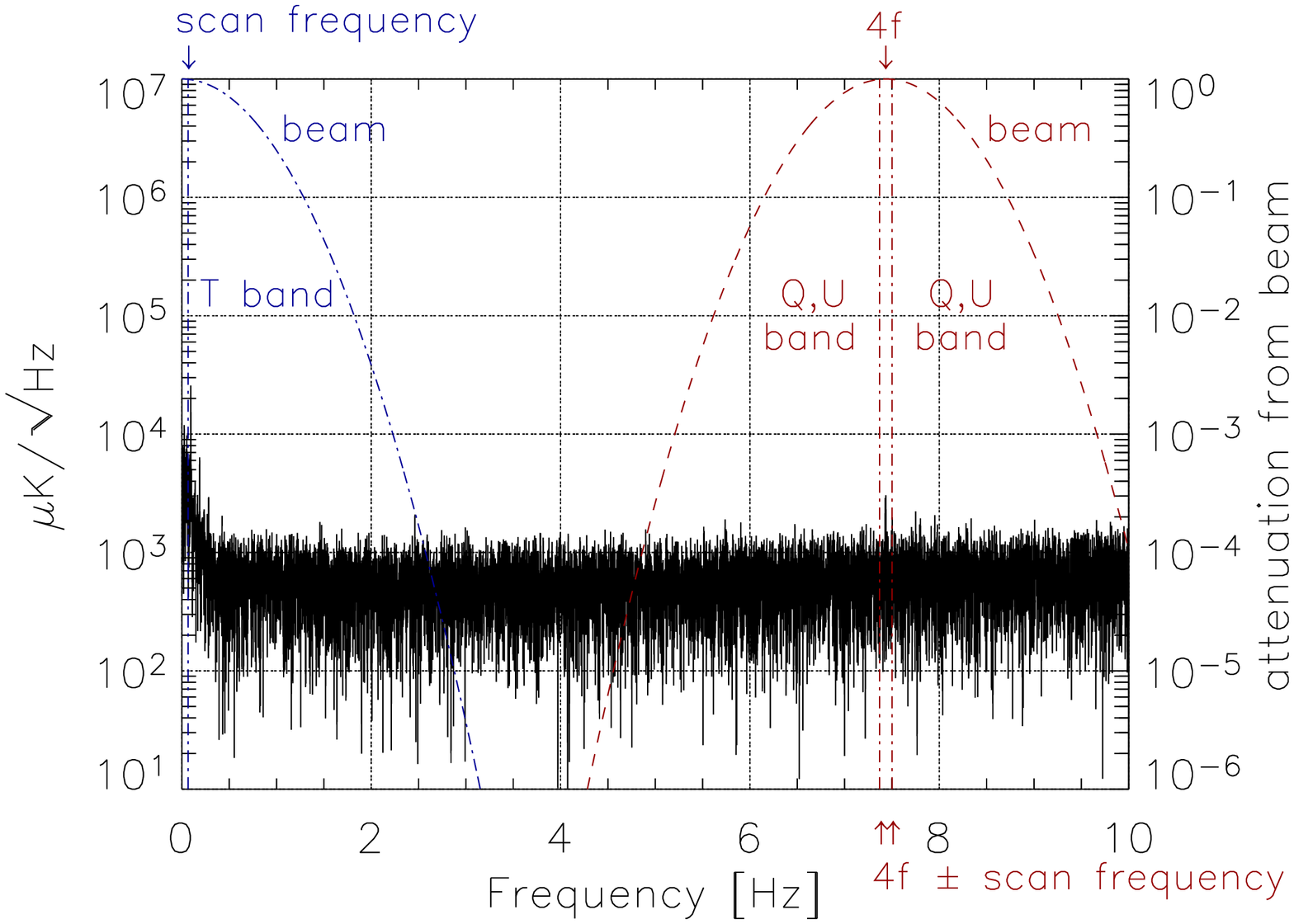} &
\includegraphics[height=2.6in,angle=90]{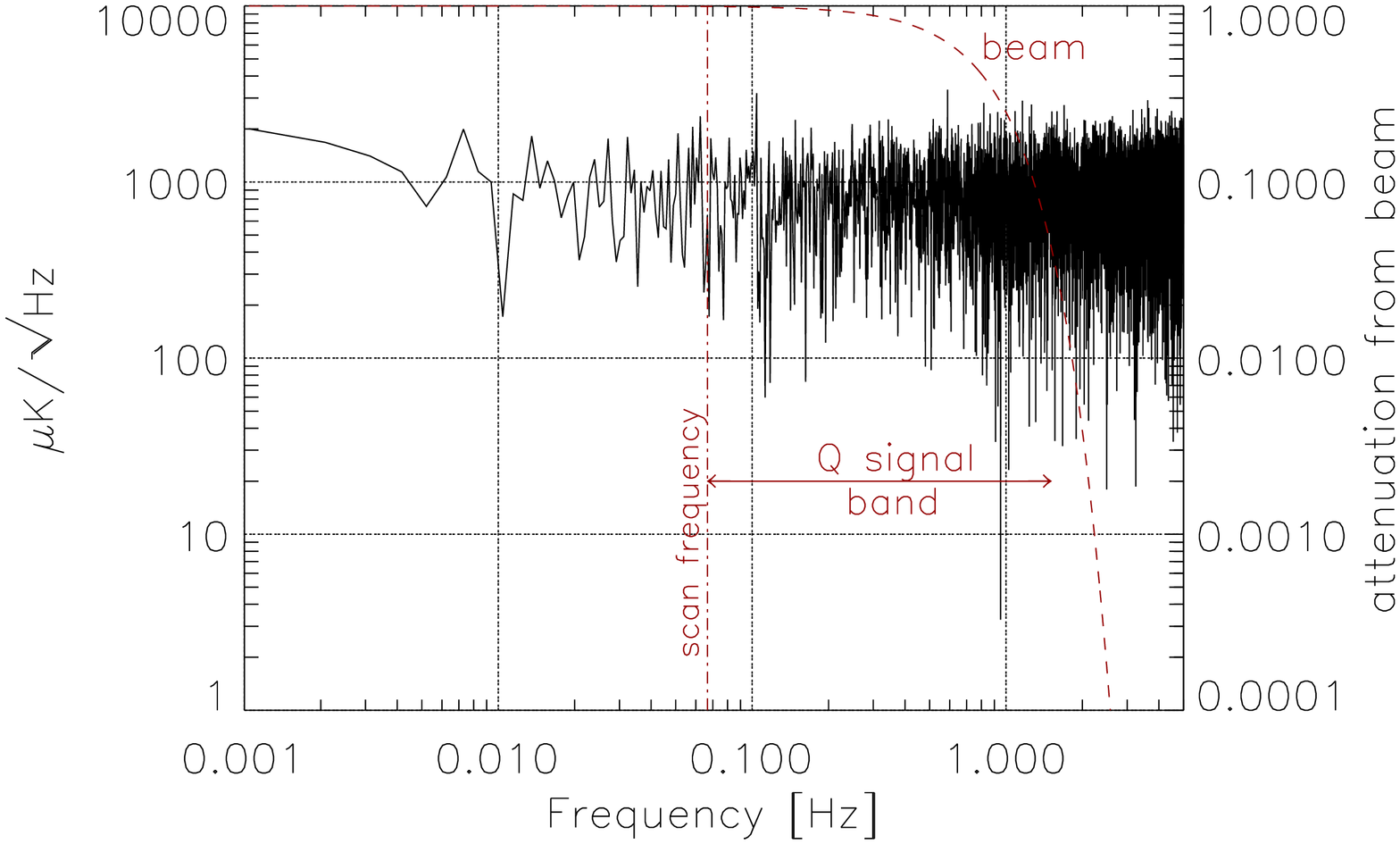} \\
(c) & (d) \\
\end{array}
$
\end{center}
\caption{\footnotesize{ $(a)$ A 16~min long sample of raw data from
one 140~GHz polarimeter plotted versus the HWP angle $\beta$.  $(b)$
Power spectrum of the raw data.  $(c)$ Power spectrum of the data
after subtraction of the \hwpss\ (TOD).  $(d)$ Power spectrum of the
demodulated TOD, which is the TOPD.  Zero-frequency in $(d)$ is
$4f_{o}$ in $(b)$ and $(c)$.  The red, dashed curves in panels $(b)$,
$(c)$ and $(d)$ mark the polarization signal bands limited by the
telescope scan frequency and the beam.  The blue, dash dot curves in
panels $(b)$ and $(c)$ mark the temperature signal band.  Some 1/$f$
noise appears in the temperature band.}}
\label{fig:TODwithHWPSS}
\end{figure*}

The raw data were dominated by the \hwpss, which came from thermal
emission from the HWP drivetrain, differential transmittance from the
HWP, and modulated instrumental polarization.  Emission from the HWP
drivetrain appeared at harmonics of $f_{o}$ with power which decreased
with frequency.  Differential transmittance contributed primarily at
$2f_{o}$, and polarized signals generated by instrumental polarization
were dominant at $4f_{o}$.

We modeled the \ss\ as a sum of $n = 8$ harmonics of the HWP rotation
frequency with amplitudes $\xi_{n}$ and phases $\chi_{n}$
\begin{equation}
\mbox{HWPSS}\left(t\right) = \sum_{n=1}^{8} \xi_{n}(t) \, \sin \left( \, n\beta(t) +
\chi_{n}(t) \, \right)
\label{eq:template_model_1}
\end{equation}
where HWPSS is the \ss.  Recall that $\beta(t)$ is the HWP rotation
angle in the instrument frame.  We used only eight harmonics because
signals at higher frequencies were blocked by the filters discussed in
Section~\ref{sec:filter_deconvolution}. Each of the three sources
listed above (thermal emission, differential transmission, and
instrumental polarization) contributed its own amplitude and phase to
produce the net amplitude $\xi$ and phase $\chi$ in each of the
harmonics. We allowed the amplitude of each of the three sources to
change linearly with time but we fixed their phase angles relative to
the instrument frame. Thus the contribution of the $n$th harmonic of
the \ss\ can be written as
\begin{eqnarray}
\mbox{HWPSS}_{n}(t) = \xi_{n}(t) \, \sin \left( \, n\beta(t) + \chi_{n}(t)  \, \right) \nonumber \\
 = \sum_{j=1}^{3} \left( A_{nj} + B_{nj} t \right) \, \sin \left( \, n\beta(t) + \phi_{nj} \, \right) 
\label{eq:template_model_2}
\end{eqnarray}
where $A_{nj}$, $B_{nj}$, and $\phi_{nj}$ are constants.
Trigonometric identities can be used to relate
Equations~\ref{eq:template_model_1}~and~\ref{eq:template_model_2} and
to rewrite the \ss\ as
\begin{eqnarray}
\mbox{HWPSS}(t) = \sum_{n=1}^{8}
  \!\!\!\!\! &( \, C_{1n} + C_{2n} t \, )& \!\!\!\! \cos \left( \, n \beta(t) \, \right) \nonumber \\
+ \!\!\!\!\! &( \, S_{1n} + S_{2n} t \, )& \!\!\!\! \sin \left( \, n \beta(t) \, \right)
\label{eq:template_model_3}
\end{eqnarray}
with appropriate relations between $\xi$, $\chi$, $A$, $B$ and the
constants $C_{1n}$, $C_{2n}$, $S_{1n}$, and $S_{2n}$. We subtracted
the \hwpss\ from the raw time domain data using fits to
Equation~\ref{eq:template_model_3} in the following way.

The 32 $C$ and $S$ coefficients were assumed to be constant within
each data segment and were estimated in each segment by an iterative
demodulation procedure.  To begin the first iteration, the data
segment was demodulated for $n = 1$ with the reference signal
$\cos(\beta)$. The coefficients $C_{11}$ and $C_{21}$ were estimated
using a linear least-squares fit to the output, ignoring the flagged
data contaminated with transients.  The component corresponding to
$C_{11}$ and $C_{21}$ was constructed and subtracted from the data.
The subtraction was repeated with the $\sin(\beta)$ reference signal,
and then for the harmonics $n = 2, ..., 8$ in order.  The process was
then iterated 25 times. After each iteration the new estimates of $C$
and $S$ were added to the old estimates.  For a typical photometer,
after four iterations of this process the difference between the data
and the \hwpss\ was observed to be indistinguishable from random noise
(see Section~\ref{sec:gaussianity_test}).

The amplitude values ranged from approximately 1.5 to 106~mK for
harmonic $n = 1$, from 30 to 250~mK for $n$ =2, and from 33 to 600~mK
for $n = 4$.  The amplitude drifts were typically 0.5\% over a 10~min
data segment and 10\% over a 3 hour time scale. The phases $\chi_n$
typically varied by less than 5~deg over the entire CMB scan. More
details about the characterization of the \hwpss\ are given in
\citet{collins_phd}.

Figure~\ref{fig:TODwithHWPSS} shows the properties of one segment of
data from one of the polarimeters before and after subtraction of the
\ss. Panel~$(c)$ shows that the power spectrum of the data after the
\ss\ was subtracted is flat in the vicinity of $4f_{0} = 7.4$~Hz and
also down to frequencies well below 1~Hz.  This white-noise level is
the nominal noise level of the instrument.

The subtraction of the \ss\ was tested using several figures of merit
which address the following questions. Are the data after subtraction
Gaussian distributed? Does the subtraction of the \ss\ remove CMB
signal from the map?  Are there residuals of the \ss\ in the map
domain?  How stable is the $4f_{0}$ component of the \ss\, and is it
well-fit by the slow time variation of the model?  We will address the
first three questions here and the others in
Section~\ref{sec:demodulation}.


\subsubsection{Gaussianity}
\label{sec:gaussianity_test}

The properties of the data after the \hwpss\ was subtracted were
tested by band-pass filtering the TOD between 0.3~and~8.9~Hz to select
the band that contains the first four Fourier modes of the \ss. The
data were decorrelated by filling the rejection band with a white
noise realization, and then averaged into 36 bins in the HWP angle
domain, each spaced by $10^{\circ}$.  The $10^{\circ}$ bin size was
selected so any residual $4f_o$ signal could be resolved.  A $\chi^2$
statistic was calculated over the 36 bins for each data segment to
search for residual \hwpss. The set of $\chi^{2}$ from each
polarimeter was compared to the $\chi^{2}$ distribution for 36
independent degrees of freedom using the KS test, and no excess was
found.


\subsubsection{Removal of CMB Signal}

We simulated the removal of the \hwpss\ with simulated raw time
streams that had the same properties as our real time streams. The
\ss, the pointing information, and the noise estimate from each
segment of data were used with simulated CMB polarization maps to
assemble a simulated raw data stream composed of the \ss, the
HWP-modulated CMB polarization signal, and a noise realization with a
power spectrum given by Equation~\ref{eq:fit.model}.  The subtraction
of the \ss\ was performed on each segment, the noise was subtracted,
and the CMB signal was recovered by demodulation (see
Section~\ref{sec:demodulation}).  The initial CMB TOPD was subtracted
from the final.  The residuals were bandpass filtered between 0.05 and
1.5~Hz, and then averaged into maps of $3^{\prime}$ pixelization using
weights set by the noise level in each segment. In the $2.3^{\circ}
\times 2.3^{\circ}$ square region used in the power spectral analysis
the residual maps had an RMS of $0.018 \pm 0.002 \mu$K in $Q$ and
$0.016 \pm 0.001 \mu$K in $U$ over five trials. These residuals were
negligible compared to the RMS of simulated CMB maps of the same
region, which was 4~$\mu$K for both $Q$ and $U$.


\subsection{Filter Deconvolution}
\label{sec:filter_deconvolution}


\begin{table*}[ht]
\begin{center}
\begin{tabular}{ccccccr@{,}l}
Photometer & $NEQ,U$ & Modulation & $\tau_A$, $\tau_B$ & $W_A$, $W_B$ & Calibration & \multicolumn{2}{c}{FWHM} \\
& $\mu \mbox{K} \sqrt{\mbox{sec}}$ & Efficiency & [ms] & [weight] &
$ \times 10^{-5}$ [V K$^{-1}$] & \multicolumn{2}{c}{[arcmin]} \\
\hline
b13 & 780  & 0.92 &  5, 27 & 0.78, 0.22 & 2.30 &  9.2 & ~10.2 \\
b14 & 2000 & 0.93 & 11, 41 & 0.55, 0.45 & 1.56 &  9.8 & ~9.9 \\
b15 & 1100 & 0.90 &  8, 29 & 0.64, 0.36 & 2.32 &  9.3 & ~10.8 \\
\hline
b23 & 780  & 0.93 &  8, 41 & 0.66, 0.34 & 2.18 & 10.0 & ~11.1 \\
b24 & 870  & 0.93 &  8, 50 & 0.28, 0.72 & 3.32 & 10.2 & ~11.3 \\
b25 & 710  & 0.93 & 10, 56 & 0.44, 0.56 & 2.91 & 10.3 & ~10.7 \\
\hline
b33 & 1000 & 0.93 &  9, 46 & 0.50, 0.50 & 2.15 &  9.4 & ~10.4 \\
b34 & 910  & 0.93 &  8, 26 & 0.39, 0.61 & 1.98 &  9.3 & ~10.6 \\
b35 & 880  & 0.94 & 12, 64 & 0.49, 0.51 & 2.34 &  9.5 & ~10.6 \\
\hline
b43 & 1200 & 0.95 &  9, 46 & 0.50, 0.50 & 1.93 &  8.6 & ~11.3 \\
b44 & 1700 & 0.93 & 11, 74 & 0.38, 0.62 & 1.90 &  9.0 & ~11.4 \\
b45 & 1100 & 0.92 & 14, 57 & 0.42, 0.58 & 2.42 &  9.8 & ~12.1 \\
\hline
receiver & 280 & 0.93 &  9, 46 & 0.50, 0.50 & 2.28 & 9.5 & ~10.9 \\
\hline
\end{tabular}
\caption{\footnotesize{Characteristics of each of the 140~GHz
    polarimeters in the array.  The row labeled ``receiver'' contains
    the inverse square sum of all $NEQ,U$ values thereby giving the
    total receiver performance.  Other elements in this row are simple
    averages.  The two bolometer time constants, $\tau_A$ and
    $\tau_B$, and their corresponding weights, $W_A$ and $W_B$, are
    differentiated with the subscripts $A$ and $B$.  Typical beam FWHM
    uncertainties from the MCMC analysis are 2\%, while typical
    calibration and modulation efficiency uncertainties are 13\% and
    2\%, respectively. The $NEQ,U$ and the calibration are consistent
    with an instrument model that includes the bolometer noise, the
    amplifier noise, the measured bolometer time constants, the
    transmission of the analyzer, and the reflectance of the HWP.}}
\label{table:results}
\end{center}
\end{table*}

The TOD were filtered in the readout electronics with high- and
low-pass Butterworth filters with edges at approximately 15~mHz and
20~Hz, respectively.  In addition, the frequency response of the
bolometers attenuated and phase-delayed the modulated $4f_{o}$ signals
at 7.4~Hz by about 35\% and $20^{\circ}$ of HWP angle, respectively.
The frequency response of the bolometers was measured in the
laboratory.  The results were modeled with two time constants and the
parameters are shown in Table~\ref{table:results}.  Alternative
analysis methods gave somewhat different parameters for the values of
the time constants and of the relative weights. These differences were
larger than the errors on the parameters within any given method.
However, calculations showed that the differences between any of the
derived filter functions in terms of amplitude and phase response at
the signal band-width near 7.4~Hz were small such that the effect on
the final results would have been negligible compared to the
calibration uncertainty or to the uncertainty on mixing between the E
and B modes due to noise.

The effects of the electronic filters and bolometer time constants
were removed from the TOD by deconvolving the complex filter
$\mathcal{F}$, where
\begin{equation}
\mathcal{F}(f) =
\mathcal{F}_{hp}\mathcal{F}_{lp}\mathcal{F}_{bolo}/|\mathcal{F}_{lp30}|.
\label{eq:filters}
\end{equation}
Here, $\mathcal{F}_{hp}$ and $\mathcal{F}_{lp}$ are the electronic
high and low-pass filters, $\mathcal{F}_{bolo}$ is the bolometer
response function, and $|\mathcal{F}_{lp30}|$ is a real,
phase-preserving software filter with a low-pass cutoff of 30~Hz.
This filter was included to suppress the spurious high frequency noise
created by the deconvolution of the bolometer response function.


\begin{figure}[h]
\begin{center}
\includegraphics[height=2.85in,angle=90]{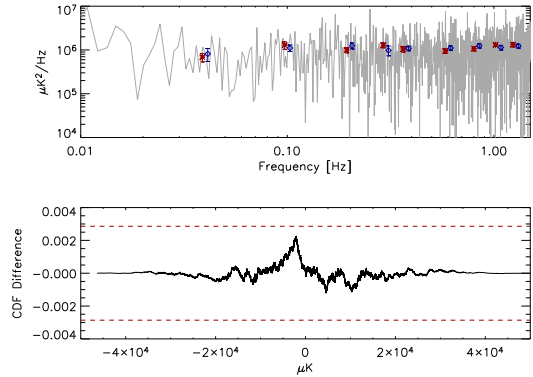}
\end{center}
\caption{\footnotesize{Top panel: The power spectrum of a 12.4~min
    long segment of TOPD for $Q$ with the binned power spectra of the
    first and second halves overplotted (red stars and blue diamonds).
    The binned power spectra are slightly offset in frequency from the
    bin center for clarity.  This segment was deemed stationary by the
    stationarity test, with $\chi^2 = 12.52$ for 9~degrees of freedom.
    The cutoff $\chi^2$ for stationarity for the 12.4~min long chunk
    length was 19.43 (see Section~\ref{sec:noise_characterization}).
    Bottom panel: the difference between the cumulative distribution
    function (CDF) of decorrelated $Q$ data from one data segment and
    the CDF of a Gaussian distribution of the same standard deviation.
    The KS statistic showed no non-Gaussianity above the 0.26$\sigma$
    level for this segment.  The red dashed lines mark the 1$\sigma$
    significance level.}}
\label{fig:StatKSplot}
\end{figure}

\subsection{Demodulation}
\label{sec:demodulation}

Modulation by the rotating HWP, combined with telescope scanning,
moves the polarized sky signals to sidebands of $4f_{o}$, as shown in
Figure~\ref{fig:TODwithHWPSS}.  To extract the $Q$ and $U$ sky signals
we used a software lock-in technique.  The TOD were demodulated with
the reference signals $-\cos(4\beta-2\alpha)/\epsilon$ and
$\sin(4\beta-2\alpha)/\epsilon$ to produce the $Q$ and $U$ TOPD,
respectively.  Demodulated polarized sky signals lie in a band defined
at the low frequency side by the $\sim$0.06~Hz telescope scan
frequency and at the high side by attenuation from the beam above
approximately 1.5~Hz, see Figure~\ref{fig:TODwithHWPSS}.  For map
making, the demodulation procedure also included a band-pass filter
with high and low-pass edges at 0.05 and 1.5~Hz, respectively, that
was used to reject any out-of-band signals.

Figure~\ref{fig:TODwithHWPSS}$(d)$ shows the power spectrum of one
segment of $Q$ TOPD.  This power spectrum is flat down to a frequency
of approximately 1~mHz.  The flatness is a measure of the degree of
stability of the $4f_{o}$ component of the \hwpss\ and the efficacy of
the subtraction method.  For 99.5\% of the TOPD from all photometers
the $1/f$ knee was below 0.06~Hz.


\subsection{Noise Characterization}
\label{sec:noise_characterization}

The TOPD were checked for stationarity by bisecting each data segment
and comparing the power spectra of the two halves. For this
comparison, the portion of the power spectra that spanned the
polarization signal band between 0.02~and~1.5~Hz was divided into nine
bins.  This choice of binning gave at least 8 frequency modes per bin
for the shortest data segments, which were approximately 2~min long.
The following $\chi^{2}$ statistic 
\begin{equation}
\chi^2 = \sum_{i=1}^9 \frac{ (P_{1i} - P_{2i})^2 }{ \sigma_{1i}^2 + \sigma_{2i}^2 }
\label{eq:comparison_chi}
\end{equation}
was used to assess the similarity of the power spectra.  Here, the sum
extends over the nine bins, and $P_i$ and $\sigma_i$ were the mean
and standard error of the power spectrum in each bin, respectively.
The distribution of this statistic for white noise time streams was
estimated with Monte-Carlo simulations of 20,000 realizations.  Seven
different segment lengths ranging from 2 to 24~min were considered.
The range of segment lengths were selected based on the maximum and
minimum lengths of the data segments used for map making. The $\chi^2$
corresponding to the probability to exceed $2 \sigma$ was stable to
3\% over this range of segment lengths and was estimated for an
arbitrary segment length by spline-interpolating the $2 \sigma$
$\chi^2$ between the seven simulated lengths.

The criterion of 2$\sigma$ stationarity was used with an algorithm
that searched for long sub-segments of stationary data.  A routine
recursively cut each data segment into two halves if the two power
spectra were dissimilar.  After bisection, an iterative routine
attempted to concatenate adjacent sub-segments whose power spectra
were not yet known to be dissimilar. The sub-segment was flagged as
non-stationary if the sub-segment length after this procedure was less
than 2~min.  Non-stationarity was mostly confined to three of the
twelve 140~GHz photometers for which approximately 11\% of the TOPD
were discarded as non-stationary. The average stationary sub-segment
length was 9~min.  The upper panel of Figure~\ref{fig:StatKSplot}
shows an example of a segment of data that was found to be stationary.

To quantify any $1/f$ noise in the TOPD the following three parameter
model
\begin{equation}
P(f) = \sigma_{w} \left[ 1 + \left( \frac{f_{knee}}{f} \right)^{a} \right] 
\label{eq:fit.model}
\end{equation}
was fit by least-squares to the binned $P_i$ of the stationary
segments with weights $\sigma_i$.  The model parameter $f_{knee}$ is
the $1/f$ knee frequency and $a$ is the spectral index of the
low-frequency noise.  The white-noise parameter $\sigma_w$ was used to
calculate the noise equivalent values of $Q$ and $U$ shown in
Table~\ref{table:results} by averaging over all stationary segments.

The decorrelated time streams $Q_d$ and $U_d$ were produced by
band-pass filtering the TOPD between 0.02 and 1.5~Hz and replacing the
out of band frequencies with white noise realizations in the frequency
domain.  The KS test was performed comparing the cumulative
distribution function (CDF) for $Q_d$ and $U_d$ to the CDF of a
Gaussian distribution of standard deviation equal to the standard
deviation of each segment.  All stationary segments passed the KS test
to better than 2$\sigma$ confidence, where the confidence level is
referred to the normal distribution.  The bottom panel of
Figure~\ref{fig:StatKSplot} shows an example of this procedure for one
segment of the data.


\section{Calibration}
\label{sec:calibration}


\subsection{Responsivity}
\label{sec:responsivity}

\begin{figure*}[ht]
\begin{center}
$
\begin{array}{cc}
\includegraphics[height=3.25in,angle=90]{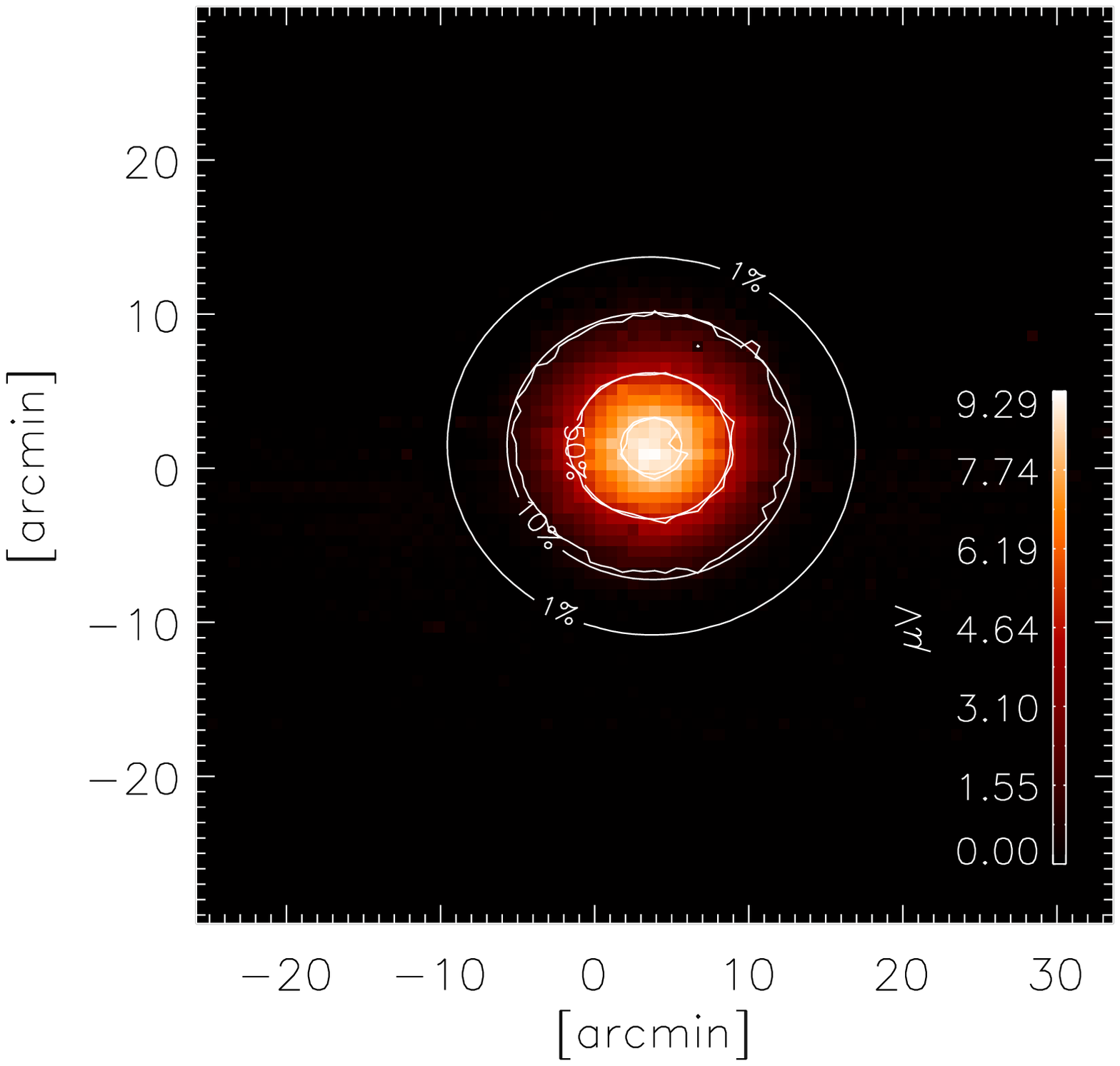} &
\includegraphics[height=3.0in,angle=90]{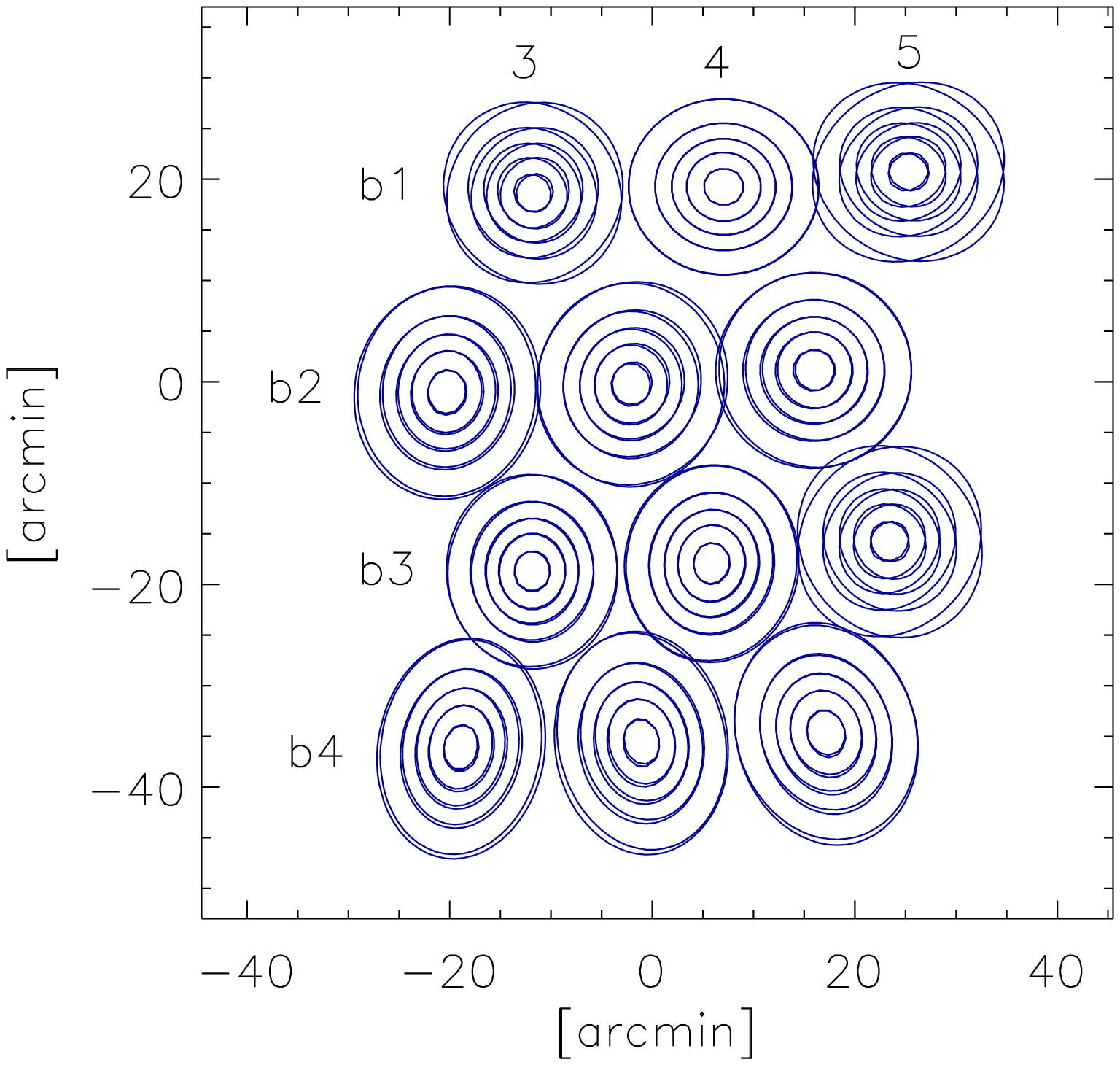} \\
(a) & (b) \\
\end{array}
$
\end{center}
\caption{\footnotesize{$(a)$ A beam map for the photometer b14 from
    the daytime observation of Jupiter.  $(b)$ Contour lines from the
    maximum likelihood Monte Carlo Markov-Chain (MCMC) fits for all
    the 140~GHz photometers.  Panel $(a)$ shows the 1, 10, 50 and 90\%
    contour lines.  The smooth lines come from the Gaussian fit and
    the jagged lines from the raw data.  Panel $(b)$ shows the 10, 30,
    50, 70 and 90\% contours from both MCMC analyses discussed in
    Section~\ref{sec:responsivity}.}}
\label{fig:beams}
\end{figure*}

MAXIPOL was calibrated with observations of Jupiter, which provided
the beam shapes, the photometer responsivities, and the beam centers
for pointing reconstruction.  In addition, the $Q$ and $U$ beams were
mapped to verify the polarization properties of the instrument.  The
time ordered data had contributions from the planet signal,
transients, the \hwpss\ and noise with a $1/f$ component.  We used the
procedure detailed in the Appendix to make a binned map of Jupiter for
both the day and nighttime scans for each of the photometers.

The beams were characterized by fitting a Gaussian to the data with a
Bayesian likelihood analysis that used a Monte Carlo Markov-Chain to
explore the likelihood function.  We used this method to mitigate the
effect of unsampled pixels in the maps, as the method explicitly
considers errors in each pixel.  We determined the following beam
parameters: the overall amplitude, the width of the beam along two
axes, the orientation angle of the ellipticity, and an overall offset,
as well as the location of the photometer in the focal plane.  In
addition to determining beam parameters from the binned map, we also
applied the analysis directly on the time ordered Jupiter data.  The
results for the two methods were consistent within the errors.  The
resulting beam widths are reported in Table~\ref{table:results}, and
the beam models are plotted in Figure~\ref{fig:beams}.  The asymmetry
of the beams was accounted for using the recipes in
\citet{proty_beams}.

The calibration $R$ reported for each photometer in
Table~\ref{table:results} was computed by integrating the
maximum-likelihood Gaussian beam
\begin{equation}
R = \frac{1}{\displaystyle T_{\mbox{\footnotesize J}} A_{\mbox{\footnotesize J}} \Gamma}~
\int \!\!\! \int V(x,y)~dx~dy
\end{equation}
where
\begin{equation}
\Gamma = \frac{\displaystyle \int S^{\prime}(\nu) \left(\partial B_{\mbox{\footnotesize RJ}}/\partial T\right)~d\nu}
{\displaystyle \int S(\nu) \left(\partial B_{\mbox{\footnotesize CMB}}/\partial T\right)~d\nu}.
\end{equation}
Here, $A_J$ is the solid angle of Jupiter, $B_{\mbox{\footnotesize
CMB}}$ is the Planck function, $B_{\mbox{\footnotesize RJ}}$ is the
Rayleigh-Jeans brightness, $V(x,y)$ is the beam map in volts and $S$
and $S^{\prime}$ are the spectral response of MAXIPOL and the
calibration apparatus, respectively.  The Rayleigh-Jeans brightness
temperature of Jupiter was taken to be $T_J = 173 \pm 9$~K
\citep{jupiter_temp}.

The temperature of the detector assembly was not regulated so the
temperature of the detectors rose with time at a rate of approximately
7~mK~hr$^{-1}$.  In order to monitor the time dependence of the
calibration, the bolometers were illuminated by a fixed-intensity
millimeter-wave lamp for 10~sec every 22~min.  The resulting bolometer
signals were found to decrease linearly with increasing bath
temperature.  The relative responsivities measured during the flashes
of the lamp were used to extend the absolute Jupiter calibration to
all times during the observations using the measured temperature of
the detector assembly.


\subsection{Polarimeter Characterization}
\label{sec:polarimeter_characterization}

\begin{figure}[ht]
\begin{center}
\includegraphics[height=3.in,angle=90]{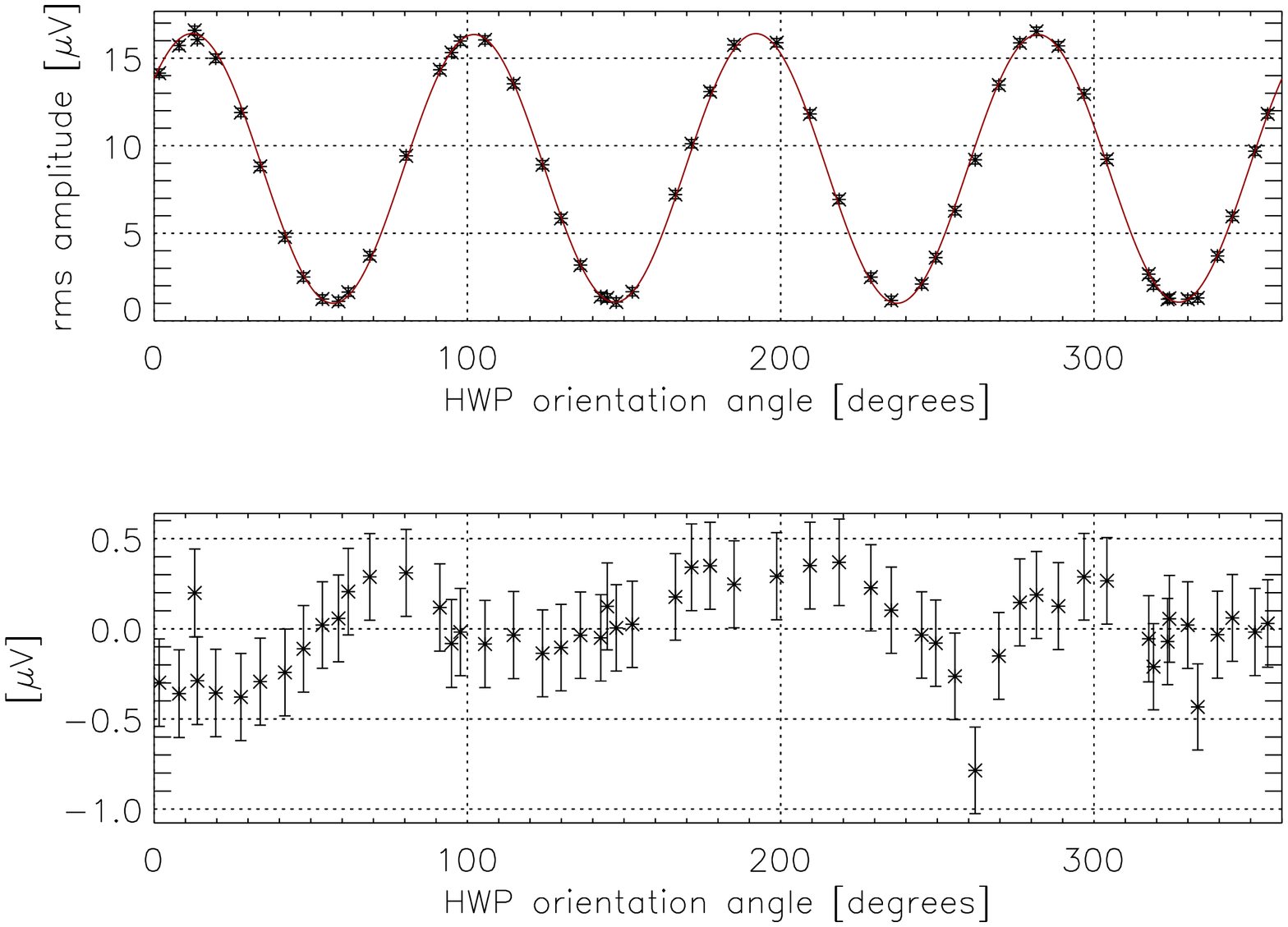}
\end{center}
\caption{\footnotesize{Data from a laboratory measurement of the
modulation efficiency.  The setup for this measurement is described in
Section~\ref{sec:polarimeter_characterization}.  The solid curve is
the best fit model given by Equation~\ref{eq:2x_4x_model}. The reduced
$\chi^{2}$ for this model is 1.03 for 42 degrees of freedom.  From the
fit parameters, we calculated $\mathcal{P}$ to be $0.88 \pm 0.03$,
which corresponds to polarimeter efficiency $\epsilon=0.94 \pm 0.02$.
The difference between these data and the best-fit model are plotted
in the bottom panel to show the goodness-of-fit and the measurement
error.}}
\label{fig:efficiency}
\end{figure}

We define the modulation efficiency $\epsilon$ as the multiplicative
factor that relates the measured data to the incident $Q$ and $U$
Stokes components, see Equation~\ref{eq:final.timestream}. It can be
shown that for monochromatic light of frequency $\nu$
\begin{equation}
\epsilon_{\nu} = g\sin^2\frac{\delta(\nu)}{2},
\label{eq:Def_epsilon}
\end{equation}
where $\delta(\nu)$ is the retardance of the HWP and $g$ is the
polarization efficiency of the wire-grid polarizer
\citep{johnson_phd,collins_phd}. We calculated a typical expected
efficiency
\begin{equation}
\epsilon_{exp} = \frac{\displaystyle \int \epsilon_{\nu} S(\nu) d\nu}{ 
    \displaystyle \int S(\nu) d\nu},  
\label{eq:Def_epsilon2}
\end{equation}
from the thickness of the HWP, the measured $g$, and a typical
spectral response as a function of frequency $S(\nu)$.  For MAXIPOL
$g$ was measured to be $0.97$~\citep{johnson_phd}.  The expected
efficiency $\epsilon_{exp}$ was 93\%, and this result agrees with the
efficiencies measured in the laboratory and listed in
Table~\ref{table:results} to within the 2\% experimental uncertainty.
The following section explains the laboratory measurement and
analysis.

A wire-grid polarizer made from the same material used to make the
polarization analyzer was mounted on the cryostat window with its
transmission axis oriented to within $2^{\circ}$ of the axis of the
analyzer. Beam filling thermal radiation from a 273~K ice bath was
chopped at approximately 6.5~Hz with a 300~K aluminum chopper blade
covered with absorbing foam (0.64~cm thick Eccosorb LS-14). In order
to avoid bolometer saturation from the brightness of the warm loads, a
4~K absorptive attenuator (1.9 cm thick Eccosorb MF110) that was
designed to transmit less than 3\% at 140~GHz was inserted into the
optical path at the intermediate focus of the telescope
\citep{eccosorb,johnson_phd}.  The HWP was stepped in $\sim5^{\circ}$
intervals and approximately 20~sec of data were collected at each HWP
orientation.  Detector drifts were removed by fitting a model
consisting of a second-degree polynomial and a single-time-constant
exponential to each data segment. The transmitted chop amplitude was
estimated by subtracting the noise RMS, which was obtained from an
independent noise-only measurement, from the data RMS.  The effect of
bolometer bath temperature drifts was mitigated by implementing a
time-dependent responsivity correction. The detector response was
linearized by applying a quadratic, amplitude dependent responsivity
correction. 

The data from one typical photometer are shown in
Figure~\ref{fig:efficiency}. The following five parameter model was
fit to the binned data
\begin{eqnarray}
m(\rho) = \!\!\!\! &A_{2f_{o}}& \!\!\!\! \cos( 2 \beta + \phi_{2f_{o}} ) \nonumber \\
       +  \!\!\!\! &A_{4f_{o}}& \!\!\!\! \cos( 4 \beta + \phi_{4f_{o}} ) + B.
\label{eq:2x_4x_model}
\end{eqnarray}
The nominal chop amplitude error per HWP orientation bin was estimated
to be the detector noise RMS.  This random error was assigned before
responsivity corrections.  The subsequent responsivity corrections
caused the magnitude of the random error to vary from bin to bin.  A
systematic error was estimated by subtracting the best fit model from
the binned data, computing the RMS of the residual and then adding
this error in quadrature with the nominal error per bin.  This
combined error was used to compute the final errors in the fit
parameters.  The experimental figure of merit
\begin{equation}
\mathcal{P} = \frac{I_{max}-I_{min}}{I_{max}+I_{min}}
\label{FOM}
\end{equation}
was calculated from the fit parameters using the equivalent
expression $\mathcal{P} = A_{4f_{o}}/B$, and the error in
$\mathcal{P}$ was propagated from the errors in $A_{4f_{o}}$ and
$B$. Using Mueller calculus it can be shown that $\mathcal{P}$ and
$\epsilon$ are related by
\begin{equation}
\epsilon = \frac{1+g^2}{g}\frac{\mathcal{P}}{1+\mathcal{P}},
\label{eq:EpsilonMeas}
\end{equation}
and the value for $\epsilon$ for each photometer is given in
Table~\ref{table:results}.  The agreement between $\epsilon$ and the
expected modulation efficiency $\epsilon_{exp}$
(Equation~\ref{eq:Def_epsilon2}) confirms that the 7\% average
depolarization for the 140~GHz photometers is accounted for by the
efficiency of the grid polarizer and the spectral response of the HWP.


\begin{figure*}[ht]
\begin{center}
$
\begin{array}{ccc}
\includegraphics[width=3.in]{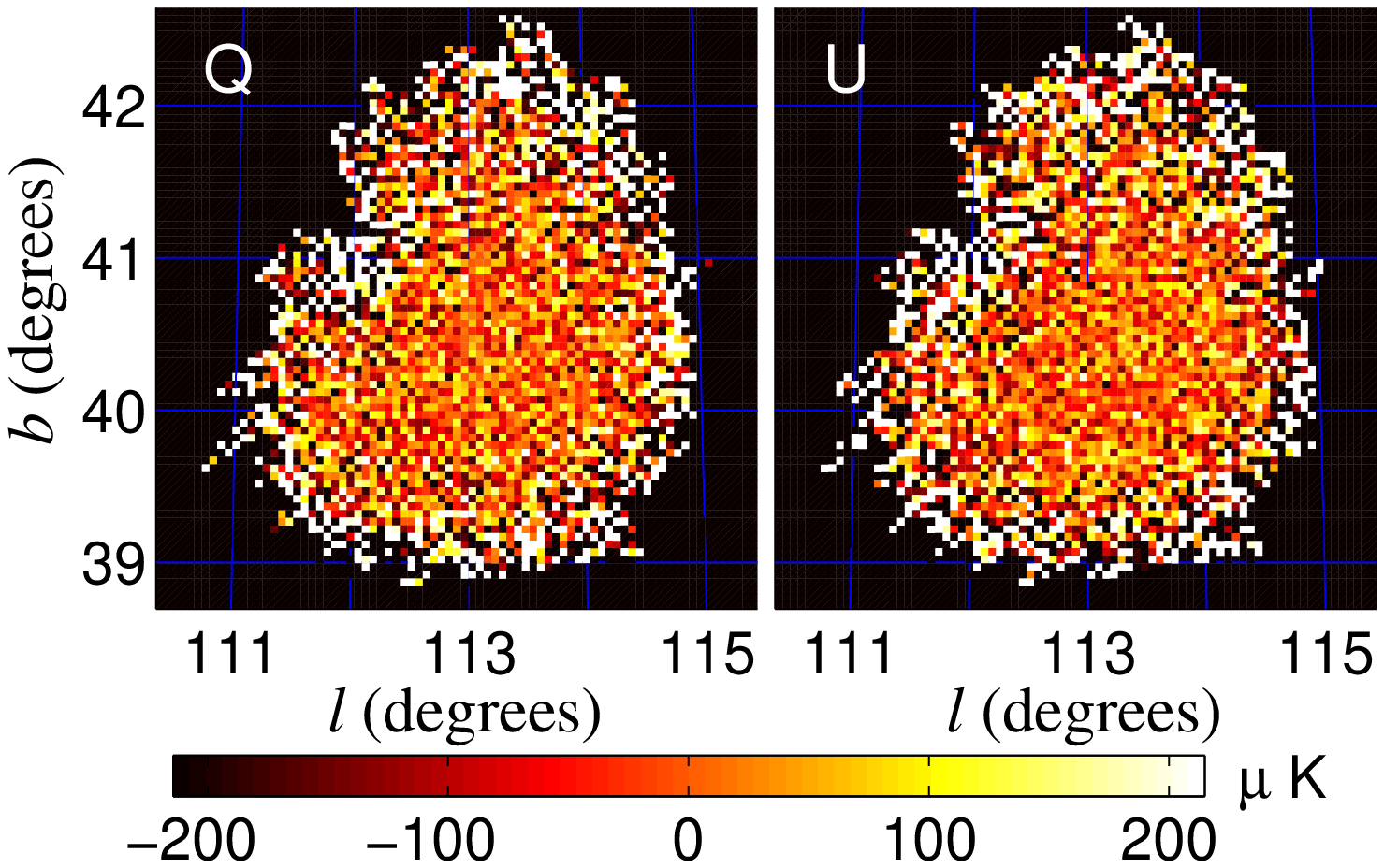} & &
\includegraphics[width=3.in]{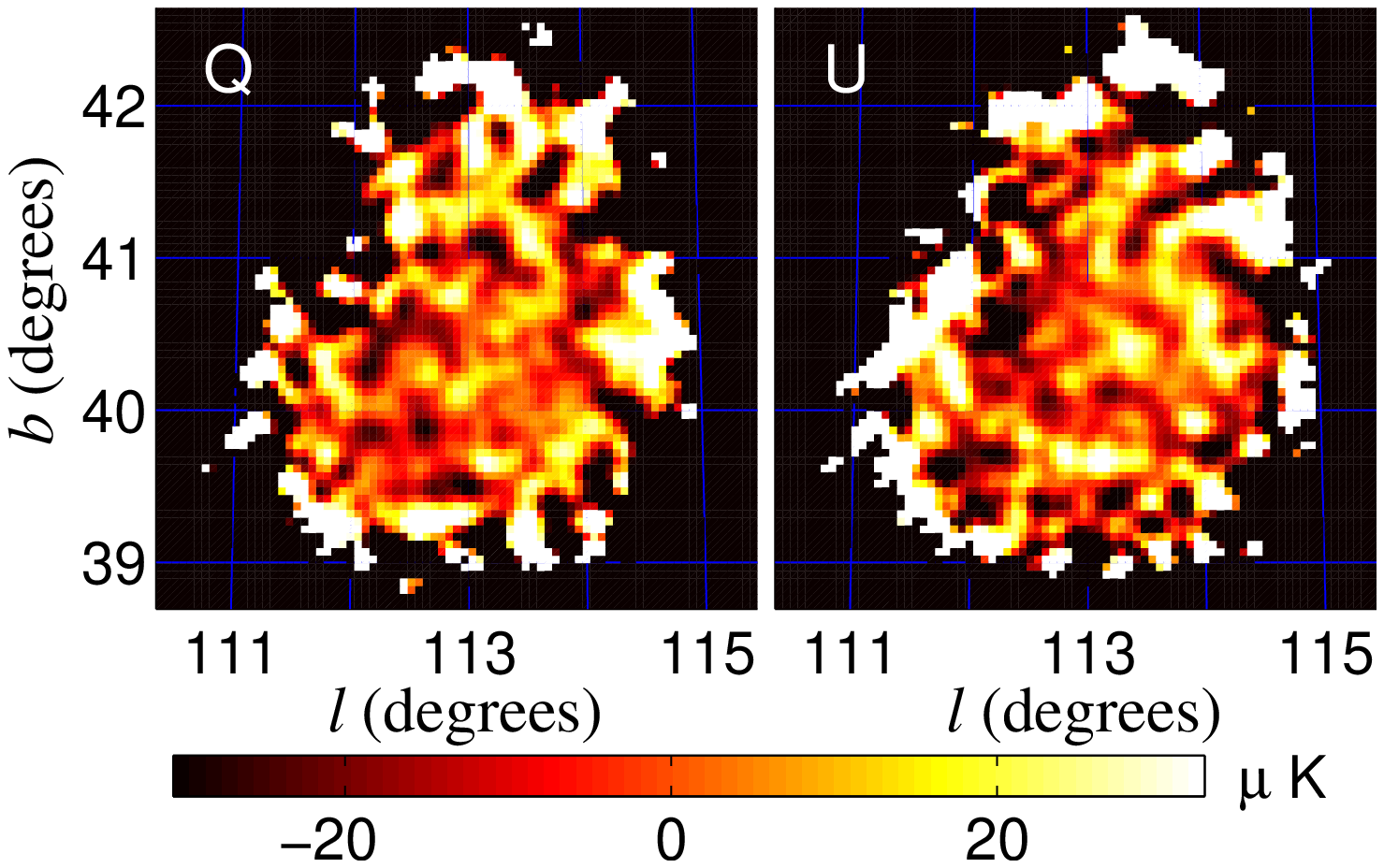} \\
\end{array}
$
\end{center}
\caption{\footnotesize{ $Q$ and $U$ maps of the MAXIPOL observation
region near \bum\ created from the raw data (left) and raw maps
convolved with a 10~arcmin FWHM Gaussian beam (right).  The pixel size
is 3~arcmin. The convolved maps are used for visual inspection only.}}
\label{fig:maps}
\end{figure*}

\section{Analysis and Results}
\label{sec:results}

We used a maximum-likelihood method to make maps of $Q$ and $U$ from
the TOPD (see Figure~\ref{fig:maps}).  The map pixel size was set to
3~arcmin.  Two approaches, one frequentist and one Bayesian, were used
to estimate the $EE$, $BB$ and $EB$ angular power spectra in three
$\ell$ bins: $\ell \leq 150$, $151 \leq \ell \leq 693$ and $\ell \geq
694$.  Given the beam size and the sky coverage we expect that only
the center bin would have any signal.  For both methods we used the
$2.3^{\circ} \times 2.3^{\circ}$ region of the maps inside the white
square in Figure~\ref{fig:hits}.  The average integration time in this
box was 117~sec for each 3~arcmin pixel.  A comprehensive description
of the analysis is given in \citet{proty}. Here we only give a
summary.

\begin{figure}[h]
\begin{center}
\includegraphics[width=3.in]{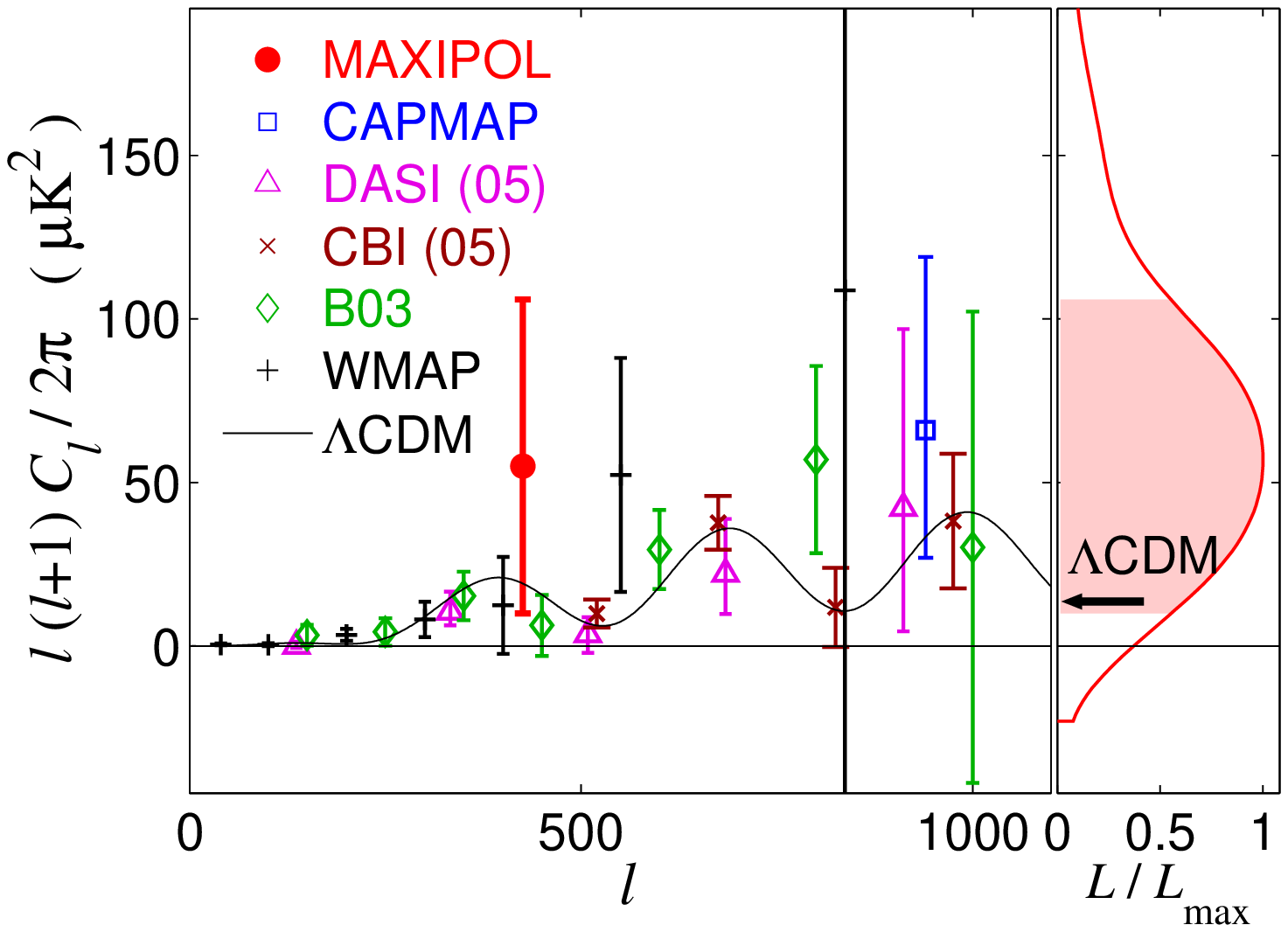}
\end{center}
\caption{\footnotesize{The MAXIPOL estimate of $EE$ power and other
    reported results (see Section~\ref{sec:introduction} for
    references).  The solid black line is the $\Lambda$CDM concordance
    model given by \citet{wmap-06cos}. The MAXIPOL result shown here
    is the 68\% probability region around the maximum of the
    likelihood function without any priors on the EB or BB
    spectra. The likelihood function is shown in the right panel.  The
    black arrow in this likelihood panel marks the expected
    $\Lambda$CDM signal for our bin, which is $14~\mu\mbox{K}^2$.
    Assuming a uniform prior the probability that the amplitude is
    positive is 96\%.  Including the calibration uncertainty does not
    change the significance of positive power and increases the error
    bars slightly, see text.}}
\label{fig:result}
\end{figure}

For the Bayesian analysis we estimated the amplitude of the angular
power spectra using a Monte-Carlo Markov Chain algorithm
\citep{mackay-information, LewisBridle} derived from the MADCAP
spectrum solver MADspec \citep{madcap}.  From a pair of 50,000 element
chains we calculated the posterior likelihoods of the amplitudes of
the power spectra in all 9 $\ell$ bins.  Estimates for the amplitude
were calculated both with and without priors which set the $EB$ and
$BB$ spectra to zero with four different functions that describe the
shape of the power spectrum within the bin.  Here we quote the results
for the amplitude of $\ell(\ell+1)C_\ell/2\pi$ after marginalization
over the eight un-interesting $\ell$ bins and without any priors on
the $EB$ and $BB$ power spectra.  Results obtained with other
assumptions are presented in \citet{proty}.

The left panel of Figure~\ref{fig:result} shows the estimated
maximum-likelihood $EE$ power and other reported results, while the
right panel shows the corresponding posterior likelihood function.
This result gives weak evidence for an $EE$ signal with a
maximum-likelihood amplitude of $55_{-45}^{+51}~\mu\mbox{K}^2$ (68\%).
The likelihood function is asymmetric and skewed positive such that if
we assume a uniform prior over all possible values of
$\ell(\ell+1)C_\ell/2\pi$ (both positive and negative) the probability
for a signal larger than zero is 96\%.  Including the calibration
uncertainty the maximum-likelihood amplitude is
$53_{-45}^{+57}~\mu\mbox{K}^2$ (68\%) with no change in the
probability of positive power. The $EE$ amplitude predicted by the
concordance $\Lambda$CDM model given by \citet{wmap-06cos} is
$14~\mu\mbox{K}^2$.  This theoretical value falls inside the 65\%
confidence region of our likelihood function around the
maximum-likelihood value.  The sharp cut-off on the negative end of
the parameter axis is a consequence of correlations that occur within
the high dimensionality of the marginalized space.  Despite this
feature, the likelihood accurately represents the result of the
experiment (see \citet{proty} for more details).

Both the 68\% confidence intervals and the significance of positive
power depend somewhat on the shape function used during the power
spectrum estimation and on whether there are prior constraints on the
$EB$ and $BB$ spectra. For example, assuming a uniform prior, the
probability that the $EE$ power spectrum amplitude is positive is 83\%
for a shape function of $1/\ell(\ell+1)$ with $EB$ and $BB$ set to
zero. The probability is 98\% for a shape function of $1/(2\ell+1)$
with no constraints on $EE$ and $EB$.

The amplitudes of the $EB$ and $BB$ power spectra assuming the
$1/\ell(\ell+1)$ shape function are $18_{-34}^{+27}$ and
$-31_{-19}^{+31}~\mu\mbox{K}^2$ (68\%), respectively, which are
consistent with $\Lambda$CDM predictions. MAXIPOL does not have the
sensitivity to detect either $BB$ or $EB$.

For the frequentist analysis we computed pseudo-band powers from the
Fourier transform of the $Q$ and $U$ maps using the flat sky
approximation. The pseudo-band powers have biases coming from the beam
convolution, the time domain processing, and the Fourier
transform. These biases were corrected, and band-power error bars were
estimated using Monte Carlo simulations and the measured beam
profiles. Results from the frequentist methods are consistent with
those from the Bayesian analysis although the error computed by the
Bayesian method is smaller, as expected.


\section{Systematic Errors}
\label{sec:systematics}

\subsection{Maps and Spectra}
\label{sec:maps_and_spectra}

We tested the $Q$ and $U$ maps used in the power spectrum analysis for
signal Gaussianity by analyzing the eigenvalue-normalized
Karhunen-Loeve coefficients \citep{gaussianity}.  If the signal is
Gaussian then these coefficients should be normally distributed.  Some
of the eigenvalues of the noise-whitened signal matrix were negative
because of the high noise and imperfectly estimated signal in those
modes.  We excluded these modes from the test, but included
noise-dominated modes.  The resulting coefficients passed the
Kolmogorov test for normality at 95\% confidence.

To check for residual systematic errors we differenced maps from the
first and second halves of the TOPD. The power spectra calculated from
these difference maps give a result that is consistent with the
absence of signal and with random noise. The maximum-likelihood value
for the $EE$ power is $-19^{+54}_{-29} \mu$K$^2$ (68\%).


\subsection{Cross-Polarization}
\label{sec:cross_pol}

We refer to ``cross-polarization'' as a leakage between $Q$ and $U$
states of polarization as quantified by the $QU$ cross term of the
appropriate Mueller matrix.  Here, we describe characterization of the
cross-polarization of the main beam.

Laboratory measurements outlined in
Section~\ref{sec:polarimeter_characterization} characterized effects
due to the photometer dependent cross-polarization of the optical
system, but without the effects of the primary mirror, combined with a
photometer independent HWP encoder offset. Both effects were corrected
in the analysis by using the photometer-dependent HWP angle $\beta$.

For the purpose of instrument characterization these two effects were
separated by assuming that the center of the focal plane has zero
cross-polarization because it is in the symmetry plane of the optical
system.  This assumption was supported by simulations of the optical
system.  The variation in the $\beta$ correction around this
zero-point was interpreted as photometer dependent cross-polarization.
The measurements show that the cross-polarization produces less than
$3.4^{\circ}$ of linear polarization rotation for any photometer.
This result agrees with ray-tracing simulations.

Simulations show that adding the primary mirror to the system should
increase the linear polarization rotation induced by
cross-polarization by less than $0.1^{\circ}$.  This additional
rotation was not corrected for during data analysis because
calculations showed the level of error in the power spectrum produced
by the effect was negligible.

Uncertainties in the angles $\alpha$ and $\beta$ (see
Section~\ref{sec:pointing}) are predominantly small systematic offsets
that are equivalent to unknown cross-polarization. The uncertainty in
$\alpha$ is $2^{\circ}$ due to the uncertainty in the orientation of
the balloon gondola during the pre-flight telescope alignment. The
uncertainty in $\beta$ is less than $2^{\circ}$ due to the uncertainty
in the alignment of the wire grid on the cryostat window during the
polarization calibration, and due to the uncertainty in the knowledge
of the bolometer time constants. According to simulation, these errors
have a negligible impact on the results.


\subsection{Instrumental Polarization}
\label{sec:inst_pol}

We refer to ``instrumental polarization'' as those effects that would
produce a detection of polarized light even if an unpolarized light is
incident on the instrument. With HWP polarimetry, polarized light that
is produced on the sky side of the HWP is modulated and gives rise to
a systematic signal at $4f_{0}$. The systematic signal can leak into
the signal bandwidth, which is at a sideband of $4f_{0}$, if it varies
with time.

In MAXIPOL several mechanisms contributed to an instrumental
polarization signal at and near $4f_{0}$. Differential reflection of
light by the primary, secondary and tertiary mirrors and by the
polypropylene vacuum window of the cryostat produced partially
polarized light in transmission. Of these, the reflection by the
window was dominant. Emission from the mirrors and from the window
produces partially polarized light because of the asymmetric nature of
the optical system. Of these emission from the primary mirror was
dominant. Unpolarized emission inside the cryostat that is diffracted
around sharp edges into the light path can produce substantial
polarization.

The instrumental polarization produced by these mechanisms was divided
into two types: a stable polarized offset and a time varying
instrumental signal that was synchronous with the sky.  The stable
offset appeared as the $4 f_{o}$ component of the \hwpss\ that
we discussed in Section~\ref{sec:hwpss_subtraction}.  The
photometer-dependent magnitude of this signal was in the range of 30
to 600~mK across the focal plane (see
Section~\ref{sec:hwpss_subtraction}). Calculations show that
instrumental polarization produced in transmission by differential
reflection off the bowed polypropylene window of the cryostat will
produce signals at the low end of this range.  The upper end of this
range is consistent with signals produced by diffraction from sharp
edges of apertures within the cold optics box. This offset was stable
on time scales much longer than data segments that were used for map
making and therefore amplitude drifts did not leak into our signal
band between 0.06 and 1.5~Hz (see Section~\ref{sec:hwpss_subtraction}
and Figure~\ref{fig:TODwithHWPSS}$(d)$). This stable signal was
rejected both during subtraction of the \hwpss\ and during
demodulation.

The time-varying instrumental signal was modulated with the intensity
pattern on the sky, thereby leaking $I \rightarrow Q, U$ and therefore
$T \rightarrow E, B$. It arose only through polarization produced in
transmission.  We measured this signal in two ways.  The level of
instrumental polarization produced by the receiver alone was
determined to be less than approximately 1\% for a typical 140~GHz
photometer in the laboratory before flight. The procedure used for
this measurement was essentially identical to the one outlined in
Section~\ref{sec:polarimeter_characterization}. However, for this
measurement the polarizer was removed from the cryostat window so the
chopped signal observed an unpolarized load. During flight, the effect
was measured more accurately with the full instrument during the
Jupiter calibration scan.  Since the level of polarization of Jupiter
is expected to be small \citep{millipol} the planet served as an
unpolarized point source.  The instrumental polarization produced in
transmission was measured by rotating the HWP during the beam-mapping
procedure.  The resulting $Q$ and $U$ beam maps of Jupiter yielded no
detectable instrumental polarization signal above about 1\% for ten of
the twelve 140~GHz photometers.  The remaining two photometers, which
were located at the edge of the focal plane, detected instrumental
polarization at the level of 4\% and 5\%.  Calculations show that this
level of instrumental polarization is plausible given a particular
alignment between the focal plane and the window to the cryostat.
Assuming conservatively that all photometers had approximately 4\%
instrumental polarization in transmission, simulations show that this
performance would only produce a spurious 3~$\mu$K$^2$ signal in our
$EE$ and $BB$ power spectra for the multipole bin of $\ell$ = 151 to
693.  Since this leakage signal is undetectable in our data, no
correction was applied during the analysis.


\subsection{Foregrounds}
\label{sec:foregrounds}

The region of the sky near \bum\ was selected for CMB observations
because contamination from dust, synchrotron and point sources is
expected to be negligible.  Polarized dust was our main concern
because synchrotron radiation is brightest at frequencies below the
MAXIPOL spectral bands, and point sources should not be polarized.
Extrapolating from 100~$\mu$m using Finkbeiner model 8 \citep{FSD99},
the unpolarized mean dust brightness over the $2.3^{\circ} \times
2.3^{\circ}$ square region used for power spectrum estimation should
be 4.1~$\mu$K with an RMS of 0.8~$\mu$K.  Archeops found a 5\%
polarized fraction for dust emission in the galactic plane
\citep{Archeops}.  The full sky WMAP 3-year data are consistent with
this polarized fraction \citep{wmap}.  Given these measurements, the
expected foreground polarization anisotropy over the \bum\ square
region should be less than 0.04~$\mu$K RMS, which is not detectable by
MAXIPOL.  A catalog search yielded no detectable radio or infrared
sources in our field \citep{smoot}.  No foreground corrections were
made during data analysis.

The data from the 420~GHz polarimeters can be used to improve the
knowledge of the level of foreground dust contamination in the \bum\
region.  However, this 420~GHz information was not used in this
analysis for two primary reasons.  First, the 420~GHz maps of Jupiter
were poorly sampled so more work is required to determine the
intensity calibration and the beam centers for pointing
reconstruction.  Second, the attenuator used for pre-flight
polarimeter characterization (see
Section~\ref{sec:polarimeter_characterization}) did not transmit
detectable amounts of 420~GHz signal, so a dedicated 420~GHz
post-flight polarimeter characterization is required to calibrate the
TOPD of this data set.


\section{Discussion}
\label{sec:discussion}

MAXIPOL was designed to be a pathfinder for bolometric CMB
polarization experiments. It is the first bolometric CMB experiment to
report results with a rapid polarization modulator.

The predicted benefits of the HWP technique have been demonstrated.
Power spectra of the time domain data from the twelve independent
polarimeters gave white noise after demodulation to frequencies well
below 50~mHz for most of the data.  A significant fraction of
demodulated data had white noise at frequencies as low as 1~mHz.  The
$Q$ and $U$ data showed no detectable systematic errors despite a
sizable \hwpss\ in the raw data. There was no need to compare noise
and responsivity data between detectors, which is necessary when using
differencing polarimeters. This made the analysis simpler.

We used the $Q$ and $U$ data to construct a map and estimate the
polarization power spectra.  The time domain noise was shown to be
Gaussian and stationary.  The noise in the map is consistent with
Gaussian random noise at a level expected given instrument noise.  The
time streams and maps were subjected to multiple tests for systematic
errors with null results.

The data give weak evidence for $EE$ power at a level that is
consistent with the prevailing cosmological model.  The $EB$ and $BB$
power spectra are consistent with zero, also as expected given
instrument noise and the cosmological model.

MAXIPOL's successful experience with a HWP will inform the design of
future, more sensitive, experiments designed to characterize B-mode
polarization of the CMB.


\section*{Acknowledgments}

We thank Danny Ball and other staff at NASA's Columbia Scientific
Ballooning Facility for their outstanding support of the MAXIPOL
program. MAXIPOL is supported by NASA Grants NAG5-12718 and NAG5-3949,
by the National Science Council and National Center for Theoretical
Science for J.H.P.~Wu, by a NASA GSRP Fellowship, an NSF IRFP and a
PPARC Postdoctoral Fellowship for B.\ R. Johnson, and by the Miller
Institute at the University of California, Berkeley for H.\ Tran.  A.\
H.\ Jaffe and J.\ Zuntz acknowledge the support of PPARC.

We are grateful for computing support from the Minnesota
Supercomputing Institute at the University of Minnesota, from the
National Energy Research Scientific Computing Center (NERSC) at the
Lawrence Berkeley National Laboratory, and from the National Center
for High-Performance Computing, Taiwan.

We thank members of the Electronics \& Data Acquisition Unit in the
Faculty of Physics at the Weizmann Institute of Science in Rehovot,
Israel for the design and construction of an on-board data recorder
unit.

We gratefully acknowledge contributions to the MAXIMA payload, which
were useful for MAXIPOL, made by V. Hristov, A.E. Lange, P. Mauskopf,
B. Netterfield, and E. Pascale.  We thank P. Oxley, D. Groom,
P. Ferreira and members of his research team for useful discussions.


\section*{Appendix: Beam Mapping}
\label{sec:appendix}

The raw data collected during the Jupiter observations contained the
following systematic errors: the \hwpss, low-frequency drifts, and
transients.  These spurious signals needed to be removed to make
accurate beam maps.  The magnitude of the Jupiter signal was similar
to the magnitude of the \hwpss, so the algorithm described here was
tailored for recovering signals that were much greater than the noise.

To remove the low-frequency drifts, we first removed a preliminary
estimate of the \hwpss.  This \ss\ was modeled in this algorithm as,
\begin{eqnarray}
{\cal H}(t) = \sum_{n=1}^{8} \left( X_{n} + Y_{n} t \right) \, 
\cos \left( \, n\beta(t) + \Theta_{n} \, \right).
\label{eq:template_model_jupiter_1}
\end{eqnarray}
The fit parameters, $X_{n}$, $Y_{n}$, and $\Theta_{n}$ were estimated
using the following two step procedure.  First, to determine the phase
of each Fourier component of the \hwpss, $\Theta_n$, the following
cosine-wave model was fit to the raw data binned in the HWP angle
domain:
\begin{equation}
{\cal H}(\beta) = \sum_{n=1}^8 Z_n \cos \left( \, n \beta + \Theta_n \, \right).
\label{eq:template_model_jupiter_2}
\end{equation}
Here, $Z_n$ and $\Theta_n$ are the fit parameters, and $\Theta_n$ in
Equation~\ref{eq:template_model_jupiter_1} is equal to $\Theta_n$ in
Equation~\ref{eq:template_model_jupiter_2}.  The bin size was set to
1~deg.  Given the HWP rotation frequency, this bin size prevented
neighboring time samples from falling into the same bin.  During the
binning procedure, low-frequency noise was rejected in each bin by
high-pass filtering the raw data in the frequency domain before
binning; transients, and planet signals were rejected by setting the
bin value equal to the mode, which was estimated by iteratively
histogramming the bin data. In the second step of estimating the \ss,
the linearly varying amplitude of each Fourier component was found by
fitting a line to the demodulated data in the time domain.  For
demodulation, the reference signal,
\begin{equation}
{\cal R}(t) = \cos \left( \, n \beta(t) + \Theta_n \, \right),
\label{eq:template_reference}
\end{equation}
was phase locked by construction because the best-fit phase, which was
output from the previous step, was used for each Fourier component.
An estimate of the \hwpss\ was then constructed using $\Theta_n$ from
step one and $X_n$ and $Y_n$ from step two.  This estimate was then
subtracted from the raw data leaving only Jupiter signal, transients,
low-frequency drifts, and noise.

Low-frequency drifts, which biased the beam maps if not subtracted,
were removed from this data by iteratively fitting and subtracting
second-degree polynomials from 24~sec long segments of data.  This
segment length was selected because it was much longer than a typical
crossing time of Jupiter through the beam.  After each iteration, the
polynomial fit was subtracted and data points greater than 3 standard
deviations away from zero were ignored in the subsequent fitting
iterations.  Given the 3 standard deviation rejection criteria, the
polynomial estimates converged after three iterations.  This masking
procedure prevented transients and the planet signal from biasing the
estimate of the low-frequency drifts.

The final drift estimate was subtracted from the {\it raw} data.  The
effects of the electronic filters and the bolometer time constants
were deconvolved using the procedure given in
Section~\ref{sec:filter_deconvolution}.  A second iteration of \hwpss\
estimation was required because the \ss\ was phase shifted by the
filter deconvolution. Transients larger than the empirically
determined maximum Jupiter signal were flagged.  The beams were then
mapped using the telescope pointing, the transient flags, and the data
with \ss\ and low-frequency drifts subtracted.  The map pixel size was
set to 0.7~arcmin to allow accurate estimation of the $B_{\ell}$,
which was used during CMB power spectrum estimation.

Biases in the beam map that were introduced by the low-frequency
drifts were significant.  To allowed for better estimation and
subtraction of these drifts, the entire process outlined above was
repeated with a Jupiter signal template removed during the
low-frequency drift estimation.  This signal template was computed by
scanning a beam model with the telescope pointing.  Here the beam
model was the two-dimensional elliptical Gaussian that best fit the
pixelized maps that were output from the first iteration of the
process.



\begin{thebibliography}{99}

\bibitem[Abroe et al.(2004)]{abroe} Abroe, M. E., et al. 2004, ApJ, 605, 607

\bibitem[Balbi et al.(2000)]{balbi} Balbi, A., et al. 2000, ApJ,
545:L1; Erratum. 2001, ApJ, 558, L145

\bibitem[Barkats et al.(2005)]{capmap} Barkats, D., et al. 2005, ApJ, 619,
2, L127

\bibitem[Benoit et al.(2004)]{Archeops} Benoit, A., et al. 2004, A~\&~A, 424,
571

\bibitem[Bock (1994)]{bockblack} Bock, J. J. 1994,
Ph.D. Thesis. University of California, Berkeley

\bibitem[Borrill et al.(2006)]{madcap} Borrill, J., et al. 2006, in
preparation

\bibitem[Clemens et al.(1990)]{millipol} Clemens, D. P., et al. 1990,
PASP, 102, 1064C

\bibitem[Collins(2006)]{collins_phd} Collins, J. S. 2006,
Ph.D. Thesis. University of California, Berkeley

\bibitem[Finkbeiner et al.(1999)]{FSD99}Finkbeiner, D. P., Davis, M. \&
Schlegel, D. J. 1999, ApJ, 524, 867

\bibitem[Griffin et al.(1986)]{jupiter_temp} Griffin, M. J., et al. 1986,
Icarus, 65, 244

\bibitem[Hagman \& Richards(1995)]{hagman} Hagmann, C. \& Richards, P. L. 1995,
Cryogenics, 35, 5, 303

\bibitem[Hanany et al.(2000)]{hanany} Hanany, S., et al. 2000, ApJ, 545, L5

\bibitem[Hinshaw et al.(2003)]{WMAPdataanalysis} Hinshaw, G., et
al. 2003, ApJS, 148, 63H

\bibitem[Jaffe et al.(1999)]{forecast} Jaffe, A. H., et al. 1999, in {\it
Microwave Foregrounds}, eds. A. de Oliveira-Costa \& M. Tegmark (San
Francisco: ASP)

\bibitem[Johnson et al.(2003)]{johnson} Johnson, B. R., et al. 2003, {New
Astronomy Reviews}, 47, 1067

\bibitem[Johnson(2004)]{johnson_phd} Johnson, B. R. 2004,
Ph.D. Thesis, University of Minnesota

\bibitem[Johnson et al.(2006)]{johnson_moriond} Johnson, B. R., et al. 2006,
in Proceedings from the Rencontres de Moriond, in press

\bibitem[Lee et al.(1998)]{lee_hardware} Lee, A. T., et al. 1998, in EC-TMR
Conference Proceedings 476, 3 K Cosmology, ed. L. Maiani,
F. Melchiorri, \& N. Vittorio (Woodbury, New York: AIP), 224, preprint
(astro-ph/9903249)

\bibitem[Lee et al.(2001)]{lee} Lee, A. T., et al. 2001, ApJ, 561, L1

\bibitem[Leitch et al.(2005)]{dasi} Leitch, E. M., et al. 2005, ApJ, 624,
10L

\bibitem[Lewis \& Bridle(2002)]{LewisBridle} Lewis, A. \& Bridle,
S. 2002, Phys. Rev. D., 66, 10

\bibitem[MacKay(2003)]{mackay-information} MacKay, D. J. C. 2003,
Information Theory, Inference, and Learning Algorithms (Cambridge, UK:
Cambridge University Press)

\bibitem[Masi et al.(2006)]{boom_inst} Masi, S., et al. 2006, A\&A, 458, 687

\bibitem[Montroy et al.(2005)]{boomerang} Montroy, T. E., et al. 2006, ApJ, 647, 813

\bibitem[Page et al.(2006)]{wmap} Page, L., et al. 2006, ApJ,
submitted, preprint (astro-ph/0603450)

\bibitem[Peterson \& Richards(1984)]{eccosorb} Peterson, J. B. \& Richards,
P. L. 1984, Int. J. Infrared Millimeterwaves, 5, 12, 1507

\bibitem[Rabii(2002)]{rabii_phd} Rabii, B. 2002,
Ph.D. Thesis. University of California, Berkeley

\bibitem[Rabii et al.(2006)]{rabii_winant} Rabii, B. et al. 2006,
Rev. Sci. Instrum., 77, 071101

\bibitem[Readhead et al.(2004)]{cbi} Readhead, A. C. S., et al. 2004,
Science, 306, 5697, 836

\bibitem[Sokasian et al.(2001)]{smoot} Sokasian, A., Gawiser, E. \&
Smoot G. F. 2001, ApJ, 562, 88

\bibitem[Spergel et al.(2006)]{wmap-06cos} Spergel, D.N., et al. 2006,
\apj submitted, preprint (astro-ph/0603449)

\bibitem[Stompor et al.(2001)]{stompor} Stompor, R., et al. 2001, ApJ, 561,
L7

\bibitem[Tinbergen(1996)]{tinbergen} Tinbergen, J. 1996,
Astronomical Polarimetry (Cambridge, UK: Cambridge University Press)

\bibitem[Winant(2003)]{winant_phd} Winant, C. D. 2003,
Ph.D. Thesis. University of California, Berkeley

\bibitem[Wu et al.(2001a)]{gaussianity} Wu, J.H.P., et al. 2001a, \prl, 87, 251303

\bibitem[Wu et al.(2001b)]{proty_beams}	Wu, J.H.P., et al. 2001b, \apjs, 132, 1

\bibitem[Wu et al.(2006)]{proty} Wu, J.H.P., et al. 2006, ApJ, in preparation (companion paper)

\bibitem[Yoon et al.(2006)]{bicep} Yoon, K. W., et al. 2006. in Millimeter
and Submillimeter Detectors and Instrumentation for Astronomy III,
Proceedings of SPIE, 6275

\bibitem[Zaldarriaga \& Seljak(1997)]{zaldarriaga97} Zaldarriaga,
M. \& Seljak, U. 1997, Phys. Rev. D., 55, 1830

\end{thebibliography}
\end{document}